\documentclass[%
 aip,
 amsmath,amssymb,
 reprint,%
]{revtex4-1}

\usepackage{graphicx}
\usepackage{dcolumn}
\usepackage{bm}

\usepackage[utf8]{inputenc}
\usepackage[T1]{fontenc}
\usepackage{mathptmx}
\usepackage{etoolbox}
\usepackage{gnuplottex}
\usepackage{comment}
\makeatletter
\def\convertto#1#2{\strip@pt\dimexpr #2*65536/\number\dimexpr 1#1}
\makeatother

\usepackage{tikz}
\usetikzlibrary{external}
\usetikzlibrary{arrows,decorations.pathmorphing}
\usepackage{graphics}
\usepackage{mwe,tikz}
\usepackage[percent]{overpic}

\usepackage{pgfplots}
\usepackage[font=small,labelfont=bf,justification=justified, format=plain]{caption}
\usepackage{subcaption} 
\usepackage{paralist}       
\usepackage{xcolor}         
\usepackage{amssymb }
\usepackage{latexsym}                  
\usepackage{keyval}                    
\usepackage{moreverb}                  
\usepackage{gnuplottex}
\usepackage{amsmath}
\usepackage{chemfig}
\definecolor{mojezluta}{RGB}{3,145,155}
\definecolor{mojecervena}{RGB}{236,80,100}
\definecolor{mojemodra}{RGB}{249,168,52}
\definecolor{1blue1}{RGB}{1,4,93}
\definecolor{2blue2}{RGB}{50,150,150}

\makeatletter
\def\@email#1#2{%
 \endgroup
 \patchcmd{\titleblock@produce}
  {\frontmatter@RRAPformat}
  {\frontmatter@RRAPformat{\produce@RRAP{*#1\href{mailto:#2}{#2}}}\frontmatter@RRAPformat}
  {}{}
}%
\makeatother
\begin{document}

\preprint{AIP/123-QED}

\title{Spectro-temporal symmetry in action-detected optical spectroscopy: highlighting excited-state dynamics in large systems}
\author{K. Charvátová}%
\affiliation{Faculty of Mathematics and Physics, Charles University}

\author{P. Malý}%
\email{pavel.maly@matfyz.cuni.cz}%
\affiliation{Faculty of Mathematics and Physics, Charles University}


\date{\today}

\begin{abstract}
Multidimensional optical spectroscopy observes transient excitation dynamics through the time evolution of spectral correlations. Its action-detected variants offer several advantages over the coherent detection and are thus becoming increasingly widespread. 
Nevertheless, a drawback of action-detected spectra is the presence of a large stationary background of so-called incoherent mixing of excitations from independent states that resembles a product of ground-state absorption spectra and obscures the excited-state signal. 
This issue is especially problematic in fluorescence-detected two-dimensional electronic spectroscopy (F-2DES) and fluorescence-detected pump--probe spectroscopy (F-PP) of extended systems, where large incoherent mixing arises from efficient exciton--exciton annihilation.
In this work, we demonstrate on the example of F-2DES and F-PP an inherent spectro-temporal symmetry of action-detected spectra, which allows general, system-independent subtraction of any stationary signals including incoherent mixing. We derive the expressions for spectra with normal and reversed time ordering of the pulses, relating these to the symmetry of the system response. As we demonstrate both analytically and numerically, the difference signal constructed from spectra with normal and reversed pulse ordering is free of incoherent mixing and highlights the excitation dynamics. We further verify the approach on the experimental F-PP spectra of a molecular squaraine heterodimer and the F-2DES spectra of the photosynthetic antenna LH2 of purple bacteria.  
The approach is generally applicable to action-detected 2DES and pump--probe spectroscopy without experimental modifications and independent of the studied system, enabling their application to large systems such as molecular complexes.   

\end{abstract}

\maketitle

\section{\label{sec:introduction}Introduction}
Action-detected nonlinear spectroscopy has seen a rapid rise in popularity in recent years, with detection of fluorescence-, \cite{wagner2005rapid, Tekavec_Lott_Marcus_2007, draeger2017rapid, karki2019before,kuhn2020interpreting, maly2020coherently, agathangelou2021phase, sanders2024expanding, mcnamee2023uncovering} photocurrent,\cite{karki2014coherent,bakulin2016ultrafast} photoions \cite{roeding2018coherent} or photoelectrons.\cite{aeschlimann2011coherent,uhl2021coherent}
The experiments are typically of the pump--probe type, with the action-detected two-dimensional spectra fully resolving the third-order four-wave mixing response, probing time-dependent correlations between excited-state transitions.\cite{jonas2003two} The main differences of the action-detected approach compared to the coherent detection are the detection against a dark background, natural implementation in a co-linear geometry, and acquisition of resonant signals only. From these follow advantages such as higher sensitivity that allows measurement at lower pulse intensities and of fragile samples. The resonant signal selection avoids signal from impurities, as well as cross-phase modulation and solvent response, providing access to early-time dynamics. Action-detected spectroscopy can be easily combined with microscopy, adding sub-micrometer spatial resolution and decreasing the probed sample volume. 
Action-detected spectroscopy has been successfully used to determine the excited-state structure and dynamics of diverse systems such as molecules in solution and in gas phase,\cite{bruder2018coherent,solowan2022direct,landmesser2023two}  dimers self-assembled and attached to DNA,\cite{lott2011conformation,marcus2023studies} molecular complexes in solution \cite{kunsel2018simulating, karki2019before} and spatially resolved in a film\cite{Goetz18} and in a membrane,\cite{tiwari2018spatially} organic photovoltaic devices\cite{bakulin2016ultrafast}, quantum dots \cite{karki2014coherent} and recently even single molecules.\cite{fersch2023single,jana2024fluorescence}

However, in extended systems such as semiconductors or molecular aggregates, action-detected spectroscopy faces a challenge in form of a large stationary background signal resembling ground-state linear absorption. The reason for this stationary background is the correlation of excitations
by a nonlinear process after the interaction between the pulses, mixing otherwise independent excitations during
the signal emission.\cite{gregoire2017incoherent,mcnamee2023uncovering} Accordingly, this unwanted signal generation has been termed "incoherent mixing" and presents a serious problem of action-detected spectroscopy, since it obscures the resolved dynamic spectral correlations. In molecular systems, the excitation incoherent mixing process is most often exciton--exciton annihilation (EEA). Recently, EEA has been shown to decrease the contrast of excited-state dynamics to about $\frac{1}{N}$ where $N$ is the number of molecules in the aggregate, severely limiting the applicability of action-detected spectroscopy.
\cite{ bolzonello2023nonlinear,javed2024photosyntheticenergytransfermissing,lopez2024photoelectrochemical} 

In this work, we investigate the relation between temporal and spectral symmetry in the action-detected spectra, on the example of F-2DES. Our key finding is that stationary, time-independent signals such as the incoherent mixing are symmetric in action-detected spectroscopy under the inversion of the time ordering of all pulses, which corresponds to an interchange of the two spectral axes in F-2DES. This provides a general way to remove the stationary signals, emphasize photo-induced dynamics, and avoid the problem of incoherent mixing.

The outline of the work is as follows. First, we introduce pulse ordering in action-detected spectroscopy and discuss the related spectral symmetries. Using the projection-slice theorem,\cite{jonas2003two,hamm2011concepts} we relate the F-2DES spectra marginals to fluorescence-detected pump--probe (F-PP) signals, recently introduced by us.\cite{maly2021fluorescence} We proceed with formulation of the subtracted signal that is sensitive to time-dependent spectral features only. 
To outline the interpretation of the signal, we present a general expression for population dynamics in terms of excitonic states and lineshapes. Comparing analytical results with a numerical calculation, we verify the properties of the difference signal. Finally, we demonstrate the subtraction on both theoretical and experimental spectra of a coupled squaraine dimer and the LH2 antenna of purple bacteria.

\section{\label{symmetry0} F-2DES spectra and their symmetry}

\begin{figure}
\includegraphics[width=0.5\textwidth]{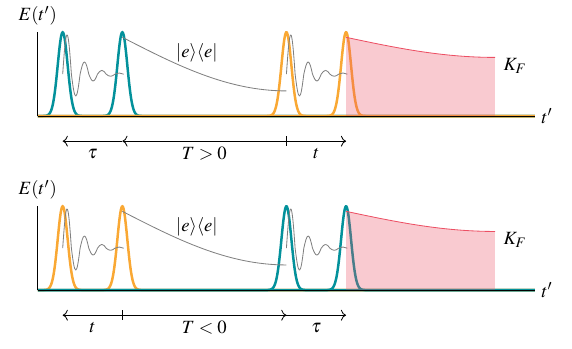}
\caption{Pulse scheme used in the F-2DES experiment. Temporal resolution is obtained by scanning time delay $T$ between pump (blue) and probe (orange) pulse pairs. The spectral resolution is obtained interferometrically by scanning the time delays $\tau$ between pump pulses and $t$ between probes. Top: standard positive time ordering of the pulses with pump followed by probe. Bottom: inverted negative time pulse ordering with probe preceding the pump. F-PP is obtained by setting $\tau=0$ in both cases.}
\label{scan}
\end{figure}
In the following, we will focus on the most popular variant of action-detected spectroscopy with fluorescence detection. However, we expect the results to be applicable to other types of action-detected spectroscopy as well. 
Fluorescence-detected two-dimensional electronic spectroscopy (F-2DES) studies nonlinear response of the system encoded in the variation of its fluorescence emission intensity, depending on the mutual time delay of pulses creating the population of excited states in the sample. The F-2DES pulse sequence is illustrated in Fig.\ref{scan}. In a typical F-2DES, the emitted fluorescence is not time- or spectrally resolved. Frequency resolution is provided by interferometric scans of two pulse pairs, the pump and the probe pairs. The temporal resolution, on the other hand, is achieved by scanning the time delay $T$ between the pump and probe pulse pair.

In standard, coherently-detected spectroscopy, the signal vanishes for negative waiting times $T$ for which the probe precedes the pump. In contrast, the population-based F-2DES features also signal for the inverted time ordering of the pulses, i.e., for $T<0$, see Fig. \ref{scan}. Upon inverting the time order of the four pulses, the role of the pump and probe pulse pairs is interchanged. One can thus expect, at least for the same pump and probe pulses, the F-2DES spectrum to be effectively transposed for each waiting time, 
\begin{equation}
\textrm{F-2DES}(\omega_t,T,\omega_\tau)\overset{?}{=}\textrm{F-2DES}(\omega_\tau,-T,\omega_t).
\label{spectro-temporal_symmetry_expectation}
\end{equation}
From this observation follows that any signal symmetric in waiting time $T$, such as the stationary incoherent mixing, will be symmetric in the F-2DES spectrum along the $\omega_\tau=\omega_t$ diagonal. This symmetry can be used to eliminate such signals.

\begin{figure}
    \centering  
\includegraphics[width=0.25\textwidth]{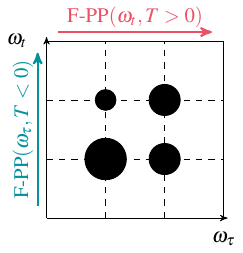}

\caption{Graphic representation of F-2DES data at a given time $T$, and the F-PP spectra as its marginals. Integration over the $\omega_\tau$ axis yields F-PP at $T>0$, while the integration over the $\omega_{t}$ axis provides F-PP at $T<0$.}
\label{2DESintegrationPP}
\end{figure}

The symmetry of F-2DES is related to the symmetry of F-PP, since F-2DES can be understood as an extension of pump--probe by the spectral resolution of the excitation, and F-PP can be obtained from F-2DES by integration over the excitation frequency axis, as outlined in Fig. \ref{2DESintegrationPP}. This relation is formalized by the so-called projection-slice theorem, \cite{hamm2011concepts,jonas2003two}  
\begin{equation}
\int \textrm{F-2DES}(\omega_t,T,\omega_\tau) d\omega_\tau= \textrm{F-PP}(T,\omega_t)
\end{equation}
The equality can be directly seen by taking the Fourier transform for $\textrm{F-2DES}(\omega_t,T,\omega_\tau)$ and realizing that its integration over $\omega_\tau$ will result in $\tau=0$, i.e., a single pump pulse. In the same way, integration over $\omega_t$ will yield $t=0$, i.e., F-PP in a negative time
\begin{equation}
\int \textrm{F-2DES}(\omega_t,T,\omega_\tau) d\omega_t= \textrm{F-PP}(-T,\omega_\tau)
\end{equation}

We now proceed to confirm the expectation of Eq. \eqref{spectro-temporal_symmetry_expectation} and its consequences by deriving the expressions for the F-2DES spectra based on a fourth-order nonlinear response to the pump and probe pulses. It is common to describe fluorescence emission as an incoherent process, by a rate dependent on the excited state kinetics.\cite{sun2024interconnection} The F-2DES spectrum can then be related to the fourth-order nonlinear response function $R^{(4)}(t_4,t_3,t_2,t_1)$,
\begin{equation}
\begin{split}
    &\textrm{F-2DES}^{(4)}(\omega_t,T,\omega_{\tau})=\iint\limits_{-\infty}^{\infty}\int\limits_0^{\infty}\cdot\cdot\cdot\int\limits_0^{\infty} R^{(4)}(t_4,t_3,t_2,t_1)E(t'-t_4) \times \\
    &\times E(t' - t_4-t_3)E(t'-t_4-t_3-t_2)E(t'-t_4-t_3-t_2-t_1) \times\\
    &\times e^{i\omega_tt} e^{i\omega_{\tau}\tau} dt_1 dt_2 dt_3 dt_4 dt' dt d\tau.
\end{split}
\label{F-2DES_convolution_general}
\end{equation}
Here, $E(t')$ is the overall electric field of all pulses in time $t'$. The response function $R^{(4)}$ can be microscopically expressed as  
\begin{equation}
\begin{split}
    &R^{(4)}(t_4,t_3,t_2,t_1)=\\
    &Tr\{\sum_i|i\rangle K_{Fi}\langle i|\mathcal{U}(t_4)[\hat{\mu},\mathcal{U}(t_3)[\hat{\mu},\mathcal{U}(t_2)[\hat{\mu},\mathcal{U}(t_1),[\hat{\mu},\hat{\rho}_0]]]] \},
\end{split}
\label{R4_definition}
\end{equation}
which describes the response of the molecular system to four interactions of the system's transition dipole moment $\mu$ with the electric field of the pulses, $\mathcal{U}(T)$ is the evolution superoperator of the system and $K_{Fi}$ corresponds to the fluorescence rate of the state $i$.

The system dynamics in F-2DES can be separated into two regions in time, one between the four interacting pulses (Fig. \ref{scan}), and the following signal emission. While some fluorescence is emitted during the pulse interaction as well, only the part that features interaction with all four pulses contributes to the nonlinear signal and gets selected by the phase cycling\cite{tan2008theory} or phase modulation.\cite{Tekavec_Lott_Marcus_2007} 
\begin{equation}
\begin{split}
    &R^{(4)}(t_4,t_3,t_2,t_1)\approx\\
    &\approx\sum_{ij}K_{Fi} U_{ij}(t_4)Tr\{|j\rangle\langle j|[\hat{\mu},\check{U}(t_3)[\hat{\mu},\check{U}(t_2)[\hat{\mu},\check{U}(t_1),[\hat{\mu},\hat{\rho}_0]]]] \}=\\
    &=\sum_{ij}K_{Fi} U_{ij}(t_4)R_j^{(3)}(t_3,t_2,t_1)
\end{split}
\label{R4approximationR3}
\end{equation}
The fourth-order action-detected response thus consists of a sum of third-order responses, weighted by their contribution to the signal during the fourth time interval. Resolving the fluorescence temporally or spectrally provides a possible additional dimension to disentangle the response contributions.\cite{Pmaly2018signatures} However, in the standard version of F-2DES, the fluorescence is integrated by a single-pixel detector. The third-order pathways are weighted by their fluorescence yields $\Phi_i$:
\begin{equation}
\begin{split}
    &\int_0^\infty dt_4 R^{(4)}(t_4,t_3,t_2,t_1)=\sum_{j}\Phi_j R_j^{(3)}(t_3,t_2,t_1)
\end{split}
\label{R4approximationR3_FLyield}
\end{equation}
Here, $R_j^{(3)}$ are the same response functions as commonly used in coherent four-wave mixing. The only difference is an additional commutator in the response from the last interaction. This produces one more excited-state-type response pathway that ends in a double excited state $ee$, while the other pathways end in the singly excited $e$ state, see Fig. \ref{2DMINUS}. According to the excitation sequence, the pathways can be divided into four groups, ground-state bleach type (GSB), stimulated emission type (SE) and excited-state absorption type contributing with one or two photons (ESA1 and ESA2, respectively), each with unique Feynman diagram shown in the first row of Fig. \ref{2DMINUS}.
In typical systems, the excitation quickly quasi-equilibrates before emitting, and the fluorescence yield of the whole excited-state manifolds can be considered state-independent.\cite{Pmaly2018signatures} Denoting the fraction of the double excitation yield to relative to that of the single excitation $\Gamma=\frac{\Phi_{ee}}{\Phi_e}$, the total signal can be simply expressed as 
\begin{equation}
   \textrm{R}_3=\textrm{R}_{\textrm{GSB}}+\textrm{R}_{\textrm{SE}}-(\Gamma-1)\cdot \textrm{R}_{\textrm{ESA}}.
\end{equation}

Since we recast in Eq.\eqref{R4approximationR3_FLyield} the fourth-order action-detected response into a linear weighted sum of third-order responses, we can use the established methodology from coherent spectroscopy, such as that by Belabas and Jonas for treating finite pulse spectra. \cite{belabas2004propagation} Since we are mainly interested in removal of the stationary incoherent mixing and emphasis of energy transfer, we follow the work of Do, Gelin and Tan.\cite{do2017simplified} As described in the Supplementary Material (SM), we derive the expressions for F-2DES spectra both in $T>0$ and $T<0$ for slowly-varying response to time-ordered pulses.
\begin{equation}
\begin{split}
    &\textrm{F-2DES}(\omega_t,T>0,\omega_{\tau})=\sum_j \Phi_j R_j^{(3)}(\omega_t,T,\omega_\tau)\times\\
    &\times I_{Pu}(\eta_1\omega_\tau-\omega_{Pu}) I_{Pr}(\eta_3\omega_t-\omega_{Pr}) e^{i\eta_1(\varphi_1-\varphi_2)+i\eta_3(\varphi_3-\varphi_4)}
\end{split}
\label{FL4_constant_positive}
\end{equation}
Here, the equations hold for a general response pathway with phase signature $(\eta_1,\eta_2,\eta_3,\eta_4)$, which can be (-1,1,1,-1) for a rephasing pathway and (1,-1,1,-1) for a nonrephasing pathway. $I_{Pu}(\omega), I_{Pr}(\omega)$ stand for the spectra of the pump and probe pulse, respectively, centered at their respective frequencies $\omega_{Pu}$, $\omega_{Pr}$. Upon inverting the pulse order, i.e., setting $T<0$ and reversing the scanning direction of $t$ and $\tau$ (see Fig. 1 in the SM for the frequency quadrants of the rephasing and non-rephasing contributions), we obtain
\begin{equation}
\begin{split}
    &\textrm{F-2DES}(\omega_t,T<0,\omega_{\tau})=\sum_j \Phi_j R_j^{(3)}(\omega_\tau,|T|,\omega_t) \times\\
    &\times I_{Pu}(-\eta_1\omega_\tau-\omega_{Pu}) 
    I_{Pr}(-\eta_3\omega_t-\omega_{Pr}) e^{i\eta_1(\varphi_1-\varphi_2)+i\eta_3(\varphi_3-\varphi_4)}
\end{split}
\label{FL4_constant_negative}
\end{equation}
Clearly, the expression for $T<0$ is identical to that for $T>0$, save for the interchanged $\omega_\tau \leftrightarrow \omega_t$ in the response function. The opposite frequency sign in the pulse spectrum simply reflects the spectral multiplication in the appropriate frequency quadrant dependent on the response type, see Fig. S1 in the SM. The anticipated  Eq. \eqref{spectro-temporal_symmetry_expectation} thus holds when the pump and probe pulses have the same spectrum. In the general case, the $T>0$ and $T<0$ spectra thus probe the symmetry of the nonlinear response function viewed through the spectral windows of the pump and probe pulses along the respective frequency axes. 

We denote the part of the response symmetric in the waiting time, i.e., with respect to swap of the first and third interval, 
\begin{equation}
    R_{\textrm{sym}}^{(4)}(0,\omega_t,T,\omega_\tau)=R_{\textrm{sym}}^{(4)}(0,\omega_\tau,T,\omega_t),
    \label{Symmetric_R}
\end{equation}
where the first zero-frequency component indicates a time-integrated fluorescence signal.

With broadband pump and probe pulses of the same spectrum covering all relevant transitions, such symmetry of the response directly corresponds to the symmetry of the 2D spectrum,
\begin{equation}
\begin{split}
    \textrm{F-2DES}_{\textrm{sym}}(\omega_t,T,\omega_\tau)&=\textrm{F-2DES}_{\textrm{sym}}(\omega_t,-T,\omega_\tau)\\
    &=\textrm{F-2DES}_{\textrm{sym}}(\omega_\tau,T,\omega_t),
    \label{Symmetric_F2DES}
\end{split}
\end{equation}
We note here that for short broadband pulses the approximation of slow dynamics in $T$ is not needed. To study the symmetry of the response with different pump and probe spectra, one can choose the region where they spectrally overlap and correct for the laser spectra. After this procedure, there is a direct correspondence between the $\omega_\tau \leftrightarrow \omega_t$ symmetry and the $T>0 \leftrightarrow T<0$ pulse-order exchange. An alternative practical solution for narrowband pulses or for F-PP is to measure the $T>0$ and $T<0$ spectra independently, by actually reversing the order of the pulses.

The spectral symmetry of the response can be leveraged to eliminate any signals that are symmetric in $T$, i.e., satisfy Eq. \eqref{Symmetric_R}, in particular those independent of $T$. To eliminate the symmetric terms, we therefore define the difference spectrum
\begin{equation}
\textrm{F-2DES}_{\textrm{diff}}(\omega_t,T,\omega_\tau)=\textrm{F-2DES}(\omega_t,T,\omega_\tau) - \textrm{F-2DES}
(\omega_\tau,T,\omega_t),
\label{F-2DES_diff_transpose}
\end{equation}
or, equivalently, using the opposite pulse ordering, 
\begin{equation}
\textrm{F-2DES}_{\textrm{diff}}(\omega_t,T,\omega_\tau)=\textrm{F-2DES}(\omega_t,T,\omega_\tau) - \textrm{F-2DES}
(\omega_t,-T,\omega_\tau).
\label{F-2DES_diff_time}
\end{equation}

A prominent signal symmetric in $T$ is the incoherent mixing contribution. Since the excitations interact during the signal emission only, the signal is a product of ground-state absorption of independent transitions in the first and third interval, satisfies the condition in Eq. \eqref{Symmetric_R} and is eliminated by the subtraction. $\textrm{F-2DES}_{\textrm{diff}}(\omega_t,T,\omega_\tau)$ is incoherent-mixing-free and includes only signals that depend on $T$ asymmetrically, such as energy transport.

The same applies for F-PP; any F-PP signals that are symmetric in waiting time must be equally present in $\textrm{F-PP}(T>0,\omega_t)$ and in $\textrm{F-PP}(T<0,\omega_t)$. 
A subtraction of these two signals, 
\begin{equation}
\textrm{F-PP}_{\textrm{diff}}(T>0,\omega_t)=\textrm{F-PP}(T>0,\omega_t) - \textrm{F-PP}(T<0,\omega_t),
\label{F-PP_diff}
\end{equation}
thus leads to a special F-PP signal sensitive to the dynamics between the interacting pulses. The $\textrm{F-PP}(T>0,\omega_t)$ and $\textrm{F-PP}(T<0,\omega_t)$ signals can be either measured separately, or derived from a single standard F-2DES spectrum measured at positive times only. 
We now proceed by discussing the properties and interpretation of these subtracted F-2DES and F-PP signals.

\section{Dynamics-sensitive difference signal}

Up to now, our formulation has been general, independent of the studied system. In the following, we will for illustration discuss the F-2DES and F-PP spectra of molecular aggregates. The primary goal of such measurements is to follow the excited-state excitonic relaxation dynamics. As previously mentioned,  this is complicated by the stationary 'incoherent mixing' background, a consequence of exciton-exciton annihilation (EEA). The excited-state dynamics is reflected in the stimulated emission (SE) signal, whose ratio to the stationary ground state bleach-type (GSB) signal, approaches $\frac{1}{N}$, for system of $N$ identical non-interacting  chromophores.\cite{bruschi2023unifying} This makes the excited state practically invisible in large systems, a situation only somewhat improved by presence of excitonic delocalization.\cite{javed2024photosyntheticenergytransfermissing} In the language of the response pathways, the incoherent mixing is expressed by a cancellation of the ESA1 and ESA2 pathways in the case of efficient annihilation (corresponding to $\Gamma=1$). The incoherently mixed signal is then captured by a GSB-type pathway with two independent excited states being probed by the pump and probe pulse pairs. Since the signal from the two molecules is 'mixed' by the EEA after the interaction with the pulses only, such signal is necessarily symmetric to the inverted order of the pulses, satisfying Eq. \eqref{Symmetric_R}. The incoherent mixing, together with all other $T$-symmetric signals, can thus be eliminated by subtraction of the spectra acquired with opposite time ordering (or transposed spectra in case of real F-2DES).

\subsection{\label{sec:level1}Signal subtraction in excitonic systems}

To demonstrate this and to establish an interpretation of the subtracted signal, we have computed the response using 4th order double-sided Feynman diagrams\cite{mukamel} with approximation of short pulses in the time domain.\cite{jonas2003two} The case of finite pulses is further discussed in section \ref{exceptions}. 
Although the pulses are spectrally broad enough to justify this approximation, we do not consider them to cover the higher excited states of the molecules. According to Kasha's rule,\cite{kasha1950characterization} excited-state absorption into these states does not lead to additional fluorescence because of their rapid relaxation. This is reflected by setting $\Gamma=1$ for these states, leading to cancellation of the ESA and ESA2 pathways and absence of these states in the spectrum. We take $\Gamma=1$ for two-exciton states as well, assuming efficient EEA as is the case in most molecular aggregates.\cite{javed2024photosyntheticenergytransfermissing} For simplicity, we assume short memory of the vibrational bath, leading to the factorization of the response into the ground-state absorptive lineshapes $G_{ig}^g(\omega)$ and excited-state stimulated-emission lineshapes $G_{ig}^e(\omega)$, and waiting time dynamics described by a propagator $\mathcal{U}(T)$. The excited state dynamics includes population transfer with rates $K_{ij}$ and possibly also a part coherently oscillating in $T$, rapidly decaying for weakly-coupled molecules. Finally, we consider a relaxation of the excited state back to the ground state with a rate $K_R$ that is much slower than the excited state dynamics. The whole signal decays with this rate, so we will omit the factor $e^{-K_R|T|}$ in the following equations for brevity. 

As shown in the SM, the standard absorptive (i.e., real rephasing plus non-rephasing) F-2DES signal at positive waiting times can then be expressed as
\begin{equation}
\begin{split}
&\textrm{F-2DES}(\omega_{t},T>0,\omega_{\tau})=\\
&-\sum_{ij}|\mu_i|^2|\mu_j|^2G^g_{ig}(\omega_{\tau})G^g_{jg}(\omega_t)\\
&-\sum_{ij}|\mu_i|^2|\mu_j|^2\mathcal{U}_{ji}(T)G^e_{jg}(\omega_t)G^g_{ig}(\omega_\tau)\\
&+R^{(4)}_\textrm{coh}(0,\omega_{t},T,\omega_{\tau})\\
\label{F-2DES_explicit_positive}
\end{split}
\end{equation}
The individual lines correspond to a correlated bleach of transitions $i$ and $j$ on the first line, absorption of state $i$, population transfer and stimulated emission of state $j$ on the second line, and oscillatory decay of an excitonic coherence on the last line. We follow the standard transient absorption convention, in which the GSB- and SE-type contributions have a negative sign (less probe-induced fluorescence in the presence of the pump pulse).  

Analogously, derived with the opposite pulse ordering, the negative waiting time F-2DES signal can be written as 
\begin{equation}
\begin{split}
&\textrm{F-2DES}(\omega_{t},T<0,\omega_{{\tau}})=\\
&-\sum_{ij}|\mu_i|^2|\mu_j|^2G^g_{jg}(\omega_{t})G^g_{ig}(\omega_{\tau})\\
&-\sum_{ij}|\mu_i|^2|\mu_j|^2\mathcal{U}_{ij}(|T|)G^g_{jg}(\omega_{t})G^e_{ig}(\omega_{\tau})\\
&+R^{(4)}_\textrm{coh}(0,\omega_{\tau},|T|,\omega_{t})\\
\label{F-2DES_explicit_negative}
\end{split}
\end{equation}
By a direct comparison with the $T>0$ signal, Eq. \eqref{F-2DES_explicit_positive}, we see that the correlated bleach term is identical (the $i\leftrightarrow j$ indices can be interchanged), while the excited-state dynamics terms have the frequencies $\omega_\tau\leftrightarrow\omega_t$ swapped. The expressions are derived in the SM to this article, including the general case of $\Gamma\neq 1$. We note here that $\Gamma \neq 1$ introduces ESA contribution to the signal, but does not change anything on the behavior of the difference signal in terms of elimination of terms symmetric in $T$.  

The fluorescence-detected pump--probe (F-PP, transient absorption) signal ($\tau$=0) can be obtained directly from the F-2DES expressions using the projection slice theorem,\cite{hamm2011concepts} by integration over $\omega_\tau$. Since the lineshapes are normalized, the F-PP expressions are obtained from F-2DES by omitting the lineshapes with $\omega_\tau$.
\begin{figure}
\includegraphics[width=0.5\textwidth]{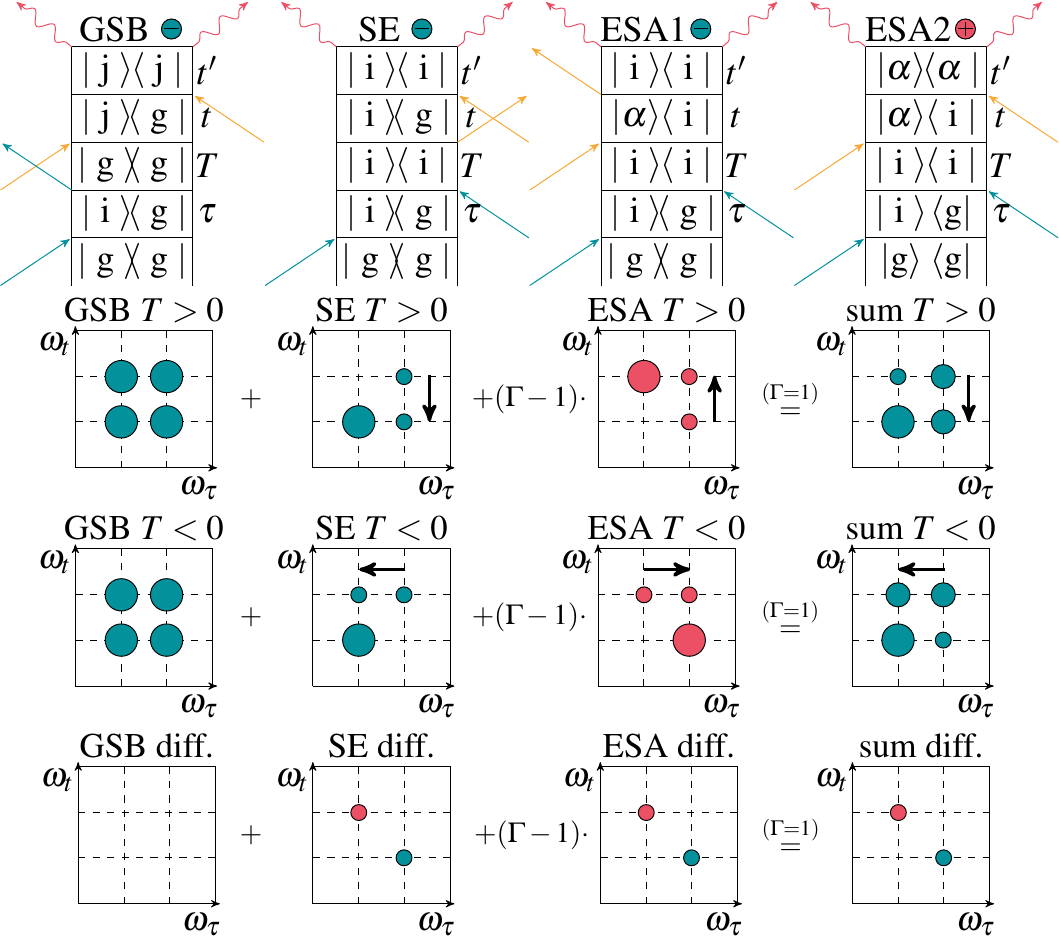}
\caption{Top row: double-sided Feynman diagrams of the response pathways contributing to the F-2DES signal of an excitonic system with a ground state $g$, excited states $i,j$ and double-excited state $\alpha$. Rephasing pathways with population dynamics are shown, for simplicty at $T=0$, separated into the GSB-type, SE-type and ESA-type. In contrast to coherently detected 2DES, the ESA pathway has the same sign as the GSB and SE, and an additional ESA2 pathway ending in a double-excited state with opposite sign is present. Blue arrows indicate interaction with the pump and probe pulses, respectively, and wavy arrows mark fluorescence emission (see Fig. \ref{scan}).  Following rows: schematic contribution of the respective pathways to the F-2DES spectra of a model heterodimer with downhill energy relaxation. Top to bottom: spectra at $T>0$ (Eq. \eqref{FL4_constant_positive} and \eqref{F-2DES_explicit_positive}), spectra at $T<0$ (Eq. \eqref{FL4_constant_negative} and \eqref{F-2DES_explicit_negative}) and the difference spectra (Eq. \eqref{F-2DES_diff_time} and \eqref{F-2DES_explicit_diff_samels}). Blue color marks contributions with a negative sign and pink with a positive, black arrow indicates energy transfer in $T$.}
\label{2DMINUS}
\end{figure}

We previously suggested subtracting the spectra in corresponding negative times to highlight excited-state relaxation in F-PP\cite{Pmaly2018signatures} and to suppress single-excitation dynamics in six-wave-mixing F-2DES.\cite{maly2020signatures} Now, we find general grounds for the subtraction based on the spectro-temporal symmetry. Subtracting the F-2DES at $T<0$ from those at $T>0$ according to Eq. \eqref{F-2DES_diff_time}, we obtain the difference spectra 
\begin{equation}
\begin{split}
&\textrm{F-2DES}_{\textrm{diff}}(\omega_{{t}},|T|,\omega_{{\tau}})=\\
&\sum_{ij}|\mu_i|^2|\mu_j|^2\times\\
&\times \left(\mathcal{U}_{ij}(|T|)G_{jg}^{g}(\omega_{t})G_{ig}^{e}(\omega_{\tau})-\mathcal{U}_{ji}(|T|)G_{jg}^{e}(\omega_{t})G_{ig}^{g}(\omega_{\tau})\right)\\
&+\left(R^{(4)}_\textrm{coh}(0,\omega_{t},|T|,\omega_{\tau})-R^{(4)}_\textrm{coh}(0,\omega_{\tau},|T|,\omega_{t})\right)
\label{F-2DES_explicit_diff}
\end{split}
\end{equation}
As expected, the constant ground-state term disappeared and only the terms reporting on excited-state dynamics remain. Same applies for the F-PP
\begin{equation}
\begin{split}
&\textrm{F-PP}_{\textrm{diff}}(|T|,\omega_{{t}})=\\
&\sum_{ij}|\mu_i|^2|\mu_j|^2\left(\mathcal{U}_{ij}(|T|)G_{jg}^{g}(\omega_{t})-\mathcal{U}_{ji}(|T|)G_{jg}^{e}(\omega_{t})\right)\\
&+\left(R^{(4)}_\textrm{coh}(0,\omega_{t},|T|,0)-R^{(4)}_\textrm{coh}(0,0,|T|,\omega_t)\right)
\label{F-PP_explicit_diff}
\end{split}
\end{equation}
The expressions are made slightly cumbersome by the presence of different lineshapes for the ground- and excited-state transitions, and in case for different lineshapes the coherent oscillations do not exactly cancel in the difference signal. In case of similar lineshapes or when their difference is removed by the spectral integration over the corresponding regions, the cancellation is complete and one obtains the simple expressions
\begin{equation}
\begin{split}
&\textrm{F-2DES}_{\textrm{diff}}(\omega_{{\tau}},|T|,\omega_{{t}})=\\
&\sum_{ij}|\mu_i|^2|\mu_j|^2\left(\mathcal{U}_{ij}(|T|)-\mathcal{U}_{ji}(|T|)\right)G_{jg}(\omega_{t})G_{ig}(\omega_{\tau}),
\label{F-2DES_explicit_diff_samels}
\end{split}
\end{equation}
(for subtraction of the oscillations see the SM). For the pump--probe hold analogously
\begin{equation}
\begin{split}
&\textrm{F-PP}_{\textrm{diff}}(|T|,\omega_{{t}})=\sum_{ij}|\mu_i|^2|\mu_j|^2\left(\mathcal{U}_{ij}(|T|)-\mathcal{U}_{ji}(|T|)\right)G_{jg}(\omega_{t}).
\label{F-PP_explicit_diff_samels}
\end{split}
\end{equation}
Clearly, with the same contribution of the ground- and excited-state lineshapes, only the population transfer dynamics remains in the subtracted signal, which features absorption at state $i$ along $\omega_\tau$ (not resolved in F-PP) and subsequent transport to state $j$ resolved spectrally along $\omega_t$. The difference signal vanishes at $T=0$ and follows the transport kinetics, with transfer into the final state $j$ increasing in magnitude with a negative sign, and transfer away from state $j$ with a positive sign. 
We note at this point that an approach that leverages below/above-diagonal symmetry of standard 2DES to extract downhill population relaxation under certain conditions was recently suggested.\cite{higgins2022leveraging}

\subsection{\label{subsec:subtraction}Difference signal of an exemplary aggregate}

We now illustrate the behavior of the difference signal on a model dimer system of two weakly coupled two-level systems with the same oscillator strengths and different transition energies, with energy transfer between the states. The F-2DES spectra of this system are shown in Fig. \ref{2DMINUS}, decomposed into the GSB, SE and ESA components. The GSB signal features two diagonal peaks at the respective transition frequencies, with cross peaks between them due to the incoherent mixing. The GSB is stationary and symmetric in $T$ and thus cancels upon subtraction. The SE features the energy transfer from the higher energy state to the lower energy one, manifesting as a decay of the upper diagonal peak and a rise of the lower cross peak with $T>0$. At negative times, the F-2DES spectra are transposed, as derived above. The subtraction eliminates the diagonal peaks, and only the population-transfer-sensitive cross peaks remain. The ESA components feature both stationary and dynamic components, but in our case they cancel because of $\Gamma=1$. The total difference signal thus resembles that of the subtracted SE and is a sensitive reporter on the population transfer dynamics. 

\begin{figure}[t!]
\begin{subfigure}[t]{0.325\textwidth}
\caption{}
\includegraphics[width=1.0\textwidth]{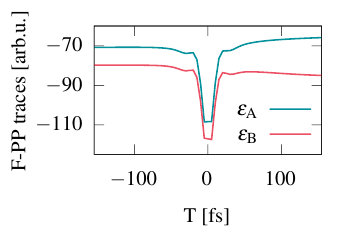}
\label{theorysubtraction:traces}
\end{subfigure}%
\begin{subfigure}[t]{0.185\textwidth}
\caption{}
\includegraphics[width=1.0\textwidth]{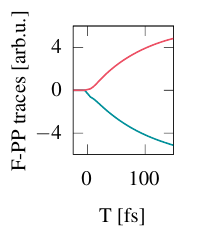}
\label{theorysubtraction:traces_diff}
\end{subfigure}
\begin{subfigure}{0.24\textwidth}
\caption{}
\includegraphics[width=0.95\textwidth]{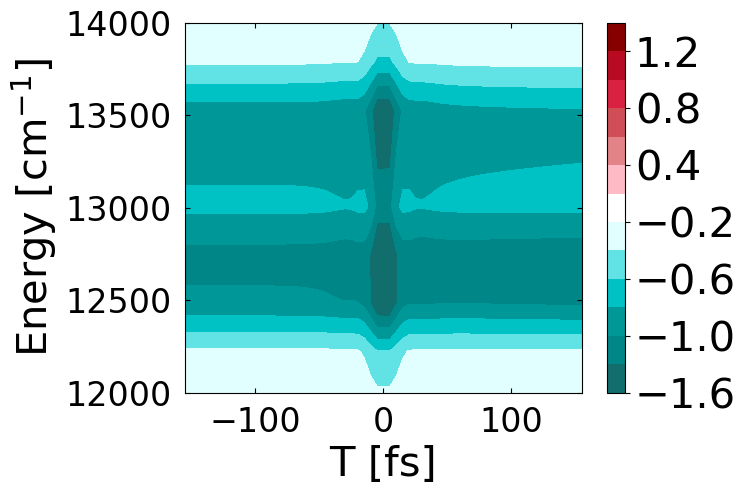}
\label{theorysubtraction:FPP}
\end{subfigure}%
\begin{subfigure}{0.24\textwidth}
\caption{}
\includegraphics[width=0.95\textwidth]{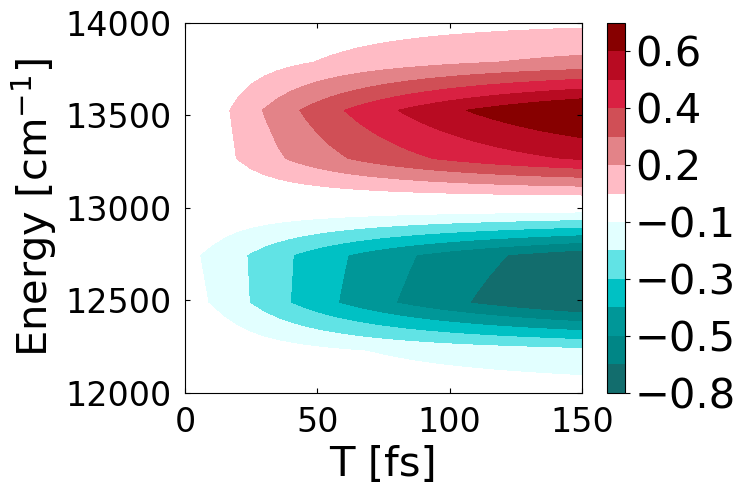}  
\label{theorysubtraction:FPP_diff}
\end{subfigure}
\caption{Numerical simulation of F-PP of a system of two groups of three molecules, with energies $\epsilon_\textrm{A}=13500\,$cm$^{-1}$ (pink color) and $\epsilon_\textrm{B}=12500\,$cm$^{-1}$ (blue color), with energy transfer with rate $K_{\textrm{BA}}=\frac{1}{100 \textrm{fs}}$ and EEA rate $K_{\textrm{annih}}=\frac{1}{20\;\textrm{fs}}$. Shown are the peak traces of F-PP (a) and difference F-PP signal (b), and $(T,\omega_t)$ transient map of the F-PP (c) and the difference signal (d). Energy transfer can be observed in (a) and (c) as the decay of the higher-energy-peak trace and the rise of the other. Signal in negative times is constant both for A and B, which corresponds to $\mu_\textrm{A}=\mu_\textrm{B}$ (see the SI for explicit form of Eqs. \eqref{F-2DES_explicit_positive}, \eqref{F-2DES_explicit_negative} for a dimer). In the unprocessed F-PP signal ((a), (c)), the visibility of dynamics is small on a large stationary incoherent mixing background. The difference signal in (b) and (d) is obtained by a subtraction of the negative-time F-PP from the positive time data. This subtraction leads to a complete removal of the constant signal background, as well as of the coherent peak around $T=0$, highlighting the excitation transfer dynamics.}
\label{theorysubtraction}
\end{figure}

The illustration in Fig. \ref{2DMINUS} is schematic and the expressions above were derived with the approximation of ultrashort pulses and Markovian dynamics. To verify that the difference signal behaves qualitatively the same with realistic pulses, we calculated the F-PP signal with 15 fs pulses based on the dynamics of a model system consisting of two groups of $(N_{\textrm{blue}},N_{\textrm{red}})$ molecules, 
with energy transfer between these two groups. We numerically integrated the corresponding master equation with Lindblad dynamics (see the SM for more detailed formulation). \cite{manvcal2020quantarhei} 
Specifically, we took $(N_{\textrm{blue}}, N_{\textrm{red}})=(3,3)$, 
with expected significant incoherent mixing and fraction $\textrm{SE}/\textrm{GSB}=1/(3+3)\approx0.17$, \cite{bolzonello2023nonlinear} energy transfer time of 100 fs and EEA time of 20 fs. In Fig. \ref{theorysubtraction:traces}, the F-PP time traces at the two absorption peaks are shown. At negative waiting times, the traces are constant, reflecting identical transition dipole moments (see second row of F-2DES spectra in Fig. \ref{2DMINUS}). Around $T=0$, the signal features a symmetric coherent peak, oscillating at the difference frequency of $1000\ \textrm{cm}^{-1}$ and rapidly decaying. As we saw in our recent work,\cite{javed2024photosyntheticenergytransfermissing} both the coherence decay and pulse overlap contribute to this $T=0$ peak. In the subtraction context, it is important that their contribution is symmetric in $T$ under these conditions.\cite{higgins2022leveraging} At $T>0$, the population transfer is visible in the higher energy peak decaying and lower energy peak rising on the large stationary incoherent mixing background. The traces thus agree well with Eqs. \eqref{F-2DES_explicit_positive}, \eqref{F-2DES_explicit_negative} integrated over the $\omega_\tau$ lineshape. In Fig. \ref{theorysubtraction}, the difference signal corresponds perfectly to Eq. \eqref{F-PP_explicit_diff_samels}: rising from zero at $T=0$, the dynamics captures the energy transfer, by both peaks increasing in magnitude, with the sign reflecting whether the energy is transferred away from them (positive sign) or into them (negative sign). As expected, both the $T=0$ peak and incoherent mixing background vanish completely upon subtraction. 

The present analysis shows a great promise of the subtraction method, indicating the utility of the spectro-temporal symmetry in F-2DES in suppressing unwanted signals and highlighting the desired dynamics. The next step is to apply the procedure to real experimental data.

\section{Exemplary applications to experimental data}

The data subtraction procedure does not require any modification of the experimental setups. Since the F-2DES (or other action-detected 2D spectra for that matter) already contain both $T>0$ and $T<0$ signals, the procedure can be applied to extract dynamics from already measured datasets. To showcase the procedure, we apply it to two systems. First, we construct the difference signal from the F-PP spectra of a coupled heterodimer, which is a relatively simple system with dynamics easily identifiable in the standard spectra as well. Then, we test the approach on energy transfer dynamics in the LH2 antenna of purple bacteria, which due to its 27 bacteriochlorophyll molecules suffers from large incoherent mixing background. In both of these systems, the EEA is highly efficient, so that $\Gamma=1$.

\begin{figure*}[t!]
\begin{subfigure}[t]{0.3\textwidth}
\caption{}
\vspace{0pt} 
    \includegraphics[width=0.9\textwidth]{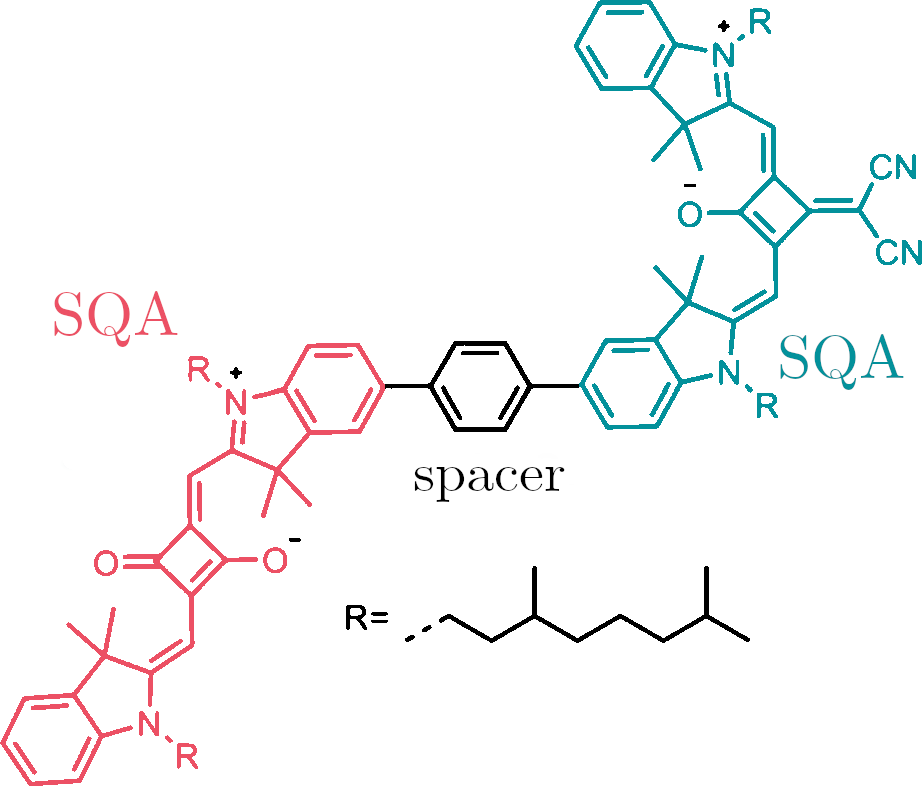}
    \label{SQ_exp_sub_structure}
\end{subfigure}%
\begin{subfigure}[t]{0.345\textwidth}
\caption{}
\vspace{0pt} 
\includegraphics[width=1.0\textwidth]{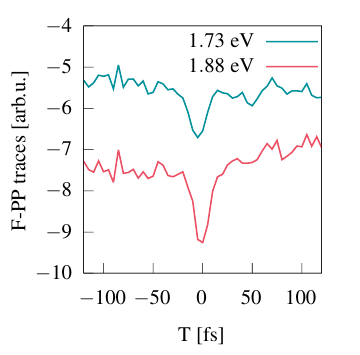}
\label{SQ_exp_sub_traces}
\end{subfigure}%
\begin{subfigure}[t]{0.345\textwidth}%
\caption{}
\vspace{0pt} 
\includegraphics[width=1.0\textwidth]{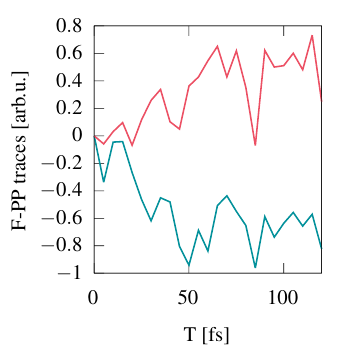}
\label{SQ_exp_sub_traces_diff}
\end{subfigure}%
\vspace{0pt}
\begin{subfigure}[t]{0.3\textwidth}
\caption{}
\vspace{0pt} 
\includegraphics[width=0.5\textwidth]{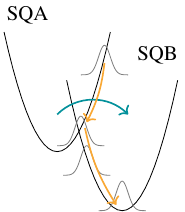}   
\label{SQ_exp_sub_relax}
\end{subfigure}
\begin{subfigure}[t]{0.345\textwidth}
\caption{}
\vspace{0pt} 
\includegraphics[width=0.9\textwidth]{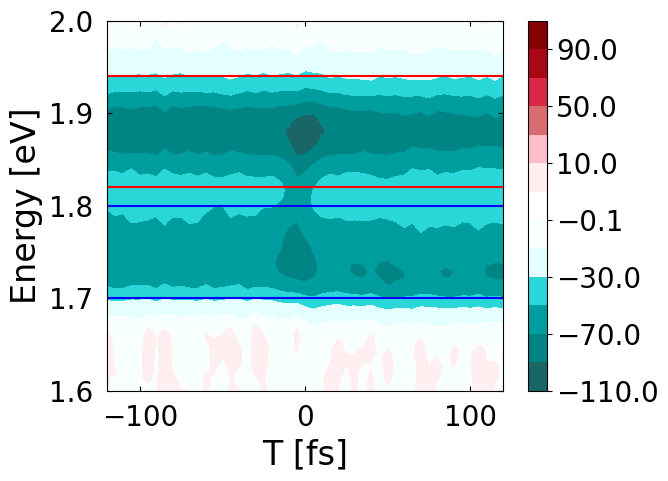}
\label{SQ_exp_sub_map}
\end{subfigure}
\begin{subfigure}[t]{0.345\textwidth}
\caption{}
\vspace{0pt} 
\includegraphics[width=0.9\textwidth]{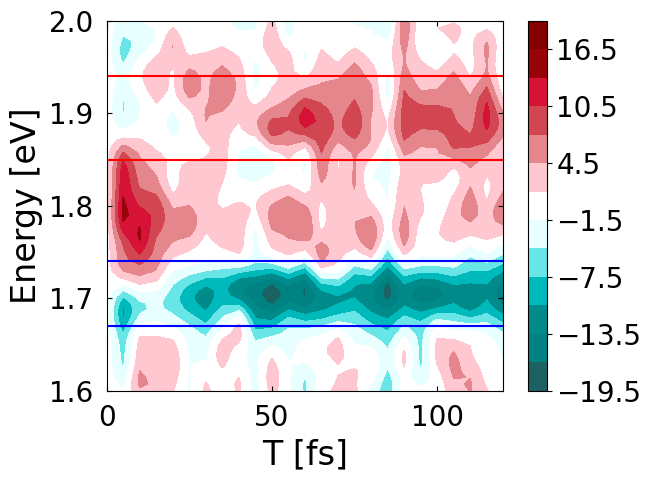}
\label{SQ_exp_sub_map_diff}
\end{subfigure}
\caption{Signal subtraction demonstrated on experimental F-PP data of squaraine heterodimer SQAB. (a) Chemcial structur of the squaraine dimer consisting of SQA and SQB molecules linked by a spacer. (b) F-PP traces of the spectral regions corresponding to the SQA (blue) and SQB (red) molecules. (c) the difference F-PP signal. (d) Diagram of excited-state vibrational relaxation and excitation transfer in the  SQAB dimer. (e) the F-PP transient map, with marked regions for traces in (b). (f) the transient map for the F-PP difference signal. Data are taken from our work in Ref. \cite{maly2021fluorescence}}.
\label{SQ_exp_sub}
\end{figure*} 

\subsection{Coupled dimer}

To demonstrate the properties of the subtracted signal, we apply the procedure to a coupled heterodimer from our previous studies.\cite{maly2020coherently,maly2021fluorescence} This particular dimer, shown in Fig. \ref{SQ_exp_sub_structure}, consists of two weakly coupled squaraine molecules, with energy transfer between their first excited states. Simultaneously, each excited state also undergoes vibrational relaxation on a time scale similar to that of energy transport (Fig. \ref{SQ_exp_sub_relax}). In Fig. \ref{SQ_exp_sub_traces} the spectral peak traces in F-PP are shown, integrated over the regions around 1.73\,eV for SQB and 1.88\,eV for SQA outlined in the F-PP map in Fig. \ref{SQ_exp_sub_map}d. Apparently, the peak traces are constant for $T<0$, in line with the theoretical expectation for similar transition dipole moments and broad spectra. At $T>0$ the higher-energy peak amplitude decreases due to energy relaxation. The lower-energy peak simultaneously receives excitation from the higher-energy SQA, but also further relaxes by excited-state reorganization marked by a dynamic Stokes shift of the SE lineshape. The subtracted F-PP data are shown in Fig. \ref{SQ_exp_sub_traces_diff} (peak traces) and Fig. \ref{SQ_exp_sub_map_diff} (transient map). The subtracted signal behaves exactly as predicted by the theoretical formulas derived above, Eq. \eqref{F-PP_explicit_diff_samels}. The higher energy peak increases to positive values signifying the decay of the negative-signed F-PP signal due to energy relaxation. The lower energy peak is, in contrast, increasing as more energy relaxes into the SQB states. In the integrated traces a faster and slower rise constants can be seen, reflecting the presence of intermolecular energy transfer and intramolecular relaxation. The SQB intramolecular relaxation can be seen by the seemingly additional positive valued feature in the subtracted F-PP map around 1.8\,eV, corresponding to the relaxing initially excited vibrational mode of SQB. 

As is clearly visible from comparison of the traces in Fig. \ref{SQ_exp_sub_traces} and \ref{SQ_exp_sub_traces_diff}, the subtraction improves the visibility of the dynamics remarkably. Since by definition the signal has to vanish at $T=0$, the contrast of dynamics is necessarily 100\%. The price to pay is the lower signal-to-noise ratio, given by the increased noise due to error propagation (factor of $\sqrt{2}$), and only the dynamic signal part remaining. Nevertheless, next to the stationary ground-state background, also the coherence peak around $T=0$ was removed by the subtraction. To summarize, the subtracted signal is somewhat noisier, but with clearly visible dynamics that is more straightforward to interpret. 

\begin{figure*}[t!]
\begin{subfigure}[t]{0.25\textwidth}
\caption{}
\vspace{0pt} 
\includegraphics[width=0.95\textwidth]{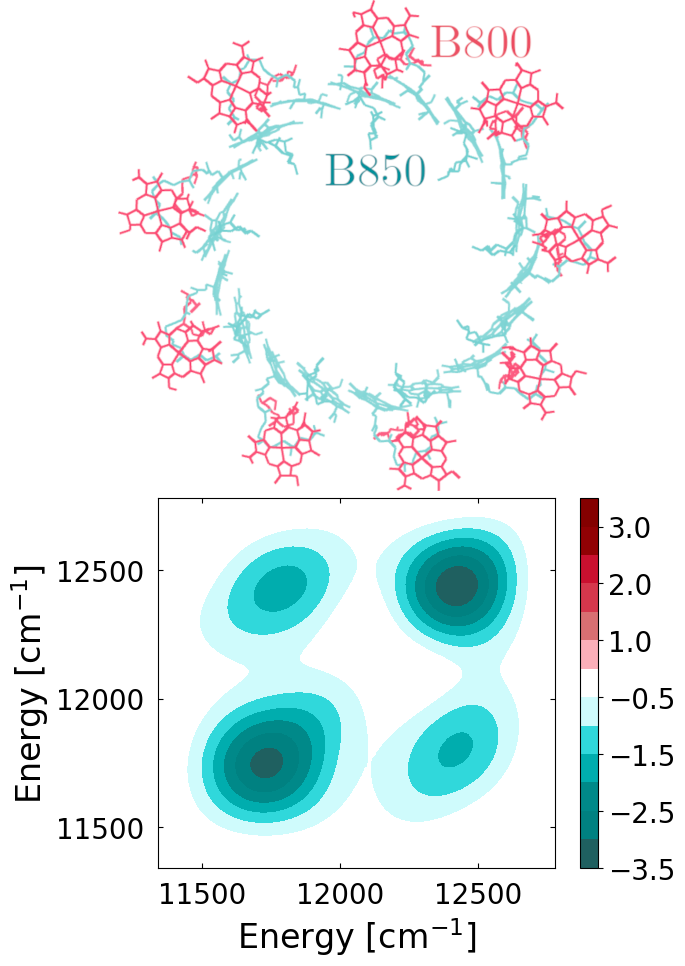}
\label{LH2_exp_sub:structure_2Dmap}
\end{subfigure}%
\begin{subfigure}[t]{0.36\textwidth}
\caption{}
\vspace{0pt} 
\includegraphics[width=1.0\textwidth]{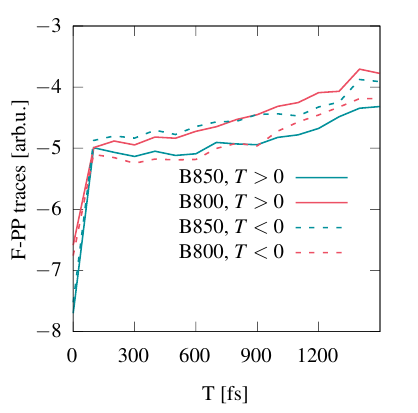}
\label{LH2_exp_sub:FPPtraces}
\end{subfigure}%
\begin{subfigure}[t]{0.36\textwidth}
\caption{}
\vspace{0pt} 
\includegraphics[width=1.0\textwidth]{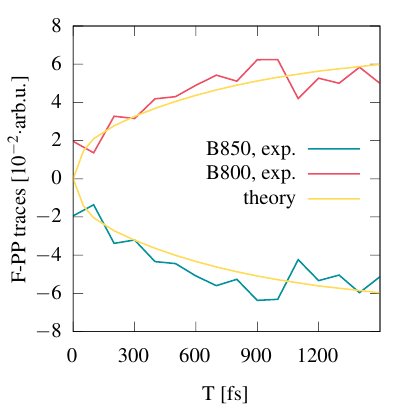}
\label{LH2_exp_sub:FPPtraces_diff}
\end{subfigure}
\caption{Demonstration and verification of the data subtraction on F-2DES data of the light-harvesting complex 2 (LH2) antenna of purple bacterium Rps. acidophila, data taken from our recent work in Ref. \cite{javed2024photosyntheticenergytransfermissing} (a,top) Structure of the LH2 antenna, with the B800 and B850 bacteriochlorophyll rings, taken from protein databank (code 2FKW \cite{cherezov2006room}). The protein scaffold is omitted for clarity. (a,bottom) F-2DES data of LH2 at $T=0\;\mbox{fs}$. (b) Traces of F-PP data, which were obtained from the F-2DES spectra by integration over $\omega_{\tau}$ for F-PP at $T>0$ and over $\omega_{t}$ for $T<0$. Spectral regions were chosen around the peaks corresponding to the B800 and B850 rings , and the spectra were integrated to obtain the presented traces (color-coded according to (a)). Solid lines are for positive times, and dashed for negative times. (c) F-PP difference signal traces (color coded by according to (a)). The difference signal was obtained by subtracting the F-PP for positive and negative waiting time and integrating the spectrum over the chosen regions to get the best signal-to-noise ratio for the final traces. Exactly the same spectral ranges were used for the theoretically calculated spectra\cite{javed2024photosyntheticenergytransfermissing} (yellow lines).}
\label{LH2_exp_sub}
\end{figure*}
\subsection{LH2 antenna of purple bacteria}

A true test of the spectro-temporal subtraction is a system large enough for the incoherent mixing to prevent reliable measurement of excited-state dynamics. We have recently encountered exactly this situation in the LH2 light-harvesting antenna of purple bacteria\cite{javed2024photosyntheticenergytransfermissing}. In the LH2 antenna of Rps. Acidophila, the excitation energy is transferred from a ring of 9 loosely-coupled bacteriochlorophylls called B800 due to its absorption wavelength, to another ring of 18 more strongly coupled pigments called B850, see Fig. \ref{LH2_exp_sub:structure_2Dmap}. The energy transfer is known from standard transient absorption to take about 0.8 ps, \cite{javed2024photosyntheticenergytransfermissing,thyrhaug2021intraband} and the EEA is highly efficient\cite{BrueggemannLH22001,javed2024photosyntheticenergytransfermissing}. In F-2DES, the  transport dynamics constitutes only about 5\% of the signal, with the rest being the incoherently mixed ground-state response of the two rings. Further relaxation dynamics to be inferred from the spectra is a rapid <100 fs energy relaxation within the B850 ring itself. 

The F-2DES data, shown in Fig. \ref{LH2_exp_sub:structure_2Dmap}, resemble the example in Fig. \ref{2DMINUS}. The B800$\rightarrow$B850 transfer should be visible in the decay of the B800-B800 diagonal peak, and in the rise of the B800-B850 cross peak. We integrated the data over $\omega_{{\tau}}$ to get F-PP for $T>0$ and over $\omega_{{t}}$ to get F-PP for $T<0$ and further over particular peaks to get traces as described in the caption to Fig. \ref{LH2_exp_sub}. Clearly, it is difficult to infer the dynamics from the $T>0$ traces only, see Fig. \ref{LH2_exp_sub:FPPtraces}. This, however, changes with the subtraction, shown in Fig. \ref{LH2_exp_sub:FPPtraces_diff}.  

The subtracted signal behaves essentially like that for a (generalized) dimer with two excitonic bands, which we have seen in section \ref{subsec:subtraction} before. At positive times, the B800-ring trace rises to positive values, reflecting energy relaxation away, and the B850 ring trace grows in negative amplitude, indicating signal rise with energy  into the B850 ring. We include in the subtracted data our previous theoretical excitonic-model calculation\cite{javed2024photosyntheticenergytransfermissing} as well, showing a good agreement. To recover the timescales present in the dynamics, we fit the calculated traces by a sum of two exponentials, obtaining rise times of $t_{1}=52.5\pm0.9\,$fs and $t_{2}=795\pm5\,$fs. These correspond perfectly to the values of 50 fs and 0.8 ps expected from theory.\cite{thyrhaug2021intraband,javed2024photosyntheticenergytransfermissing} The subtraction is thus clearly capable of removing the incoherent mixing and highlighting the excited-state dynamics even in large coupled molecular systems.

\section{Discussion}

\subsection{Practical considerations}

We demonstrated in this article a way of leveraging the spectro-temporal symmetry of stationary signals for their removal by subtraction of negative-waiting time signals from the positive-waiting-time data. In real F-2DES, The $T<0$ and $T>0$ signals are related by a spectral transposition. For similar pump and probe pulses, this allows carrying out the subtraction from a single standard F-2DES dataset. In practice, such full 2D spectrum subtraction is very sensitive to lineshape artifacts such as phase twists due to imperfect phase referencing. Furthermore, since the subtraction removes the fully-diagonal symmetric features anyway, most of the the relevant information is in the F-PP spectra, which can be obtained as marginals of the full F-2DES dataset. Furthermore, the measurement of full F-2DES is not necessary and the significantly simpler acquisition of F-PP suffices. In this case, the $T<0$ waiting times must be acquired separately. This is, however, a relatively small price to pay for the 100\% contrast of excited state dynamics. The separate measurement of $T<0$ and $T>0$ signals is necessary for non-overlapping pump and probe spectra, both in F-2DES and in F-PP.  Furthermore, the subtraction of the negative-time signal has other advantages, such as the elimination of not only all constant signals, but some coherent oscillatory signals around $T=0$ and symmetric pulse-overlap artifacts as well. The subtraction suppresses a large part of the signal and somewhat increases the noise, leading to a decreased signal to noise ratio. It can thus still happen that the dynamics in the difference signal will end up being invisible within the experimental noise. The same can be, however, expected from any method of signal suppression.

\subsection{\label{exceptions}Different pump and probe pulses}

The negative-time subtraction for removal of incoherent mixing works with any pump and probe pulses, as formulated in section \ref{symmetry0}. The presented examples on excitonic systems were, however, discussed in the impulsive limit, and the experimental spectra obtained with broadband near-identical pump and probe pulses. 
For different pump and probe spectra, the F-2DES spectra become asymmetric, and the symmetry of the response function can be studied directly from their symmetry only after a correction to the laser spectra and in the spectral overlap region. Furthermore, the correction based on Eqs. \eqref{FL4_constant_positive}, \eqref{FL4_constant_negative} works for slowly-varying signals in $T$ only. For contributions that rapidly change in $T$ on timescale comparable with pulse length, such as oscillating vibrational and vibronic wavepackets, the influence of the pulse spectra is more complex.\cite{hamm2011concepts,do2017simplified} A solution that works both in this case and in the case of spectrally distinct pump and probe pulses is to acquire the $T>0$ and $T<0$ signals independently, actually reversing the order of the pulses as shown in Fig. \ref{scan}. One can then always construct the difference signal using Eq. \eqref{F-2DES_diff_time} for F-2DES or Eq. \eqref{F-PP_diff} for F-PP. Such-constructed difference signals are incoherent mixing-free and reflect only the dynamic part of the response. The interpretation of these subtracted signals in terms of vibronic wavepackets deserves a further discussion, which is, however, beyond the scope of this work. 
The main point we wish to make here is that the subtraction process is capable of removing constant background completely even for different pump and probe pulses with realistic spectra.

\subsection{Other methods of eliminates incoherent mixing}

Ever since the formulation of its origin, the options to eliminate the incoherent mixing have been investigated.\cite{kalaee2019differentiation} The attempt to identify unique incoherent mixing phase signatures has been unsuccessful.\cite{bargigia2022identifying} A recent work suggested using a special polarization scheme to cancel the incoherent mixing contributions from isotropically oriented samples. \cite{faitz2024spectrometer} However, the polarization removes the isotropic incoherent mixing part only, which is insufficient in large molecular systems with fixed orientations, such as light-harvesting complexes. Yet another option to suppress incoherent mixing is short-time-gating or in general time resolution of the fluorescence emission, selecting the fluorescence before the mixing takes place.\cite{Pmaly2018signatures,bruschi2022simulating} This is, however, impractical due to the typically short EEA timescale that requires fast gating, leading to loss of signal intensity. 

\section{Conclusions}

We have demonstrated on the example of fluorescence-detected  2D spectroscopy a general property of action-detected 2D spectra that contributions symmetric in the waiting time, probed by a reverse ordering of the excitation pulses, are symmetric in the 2D correlation spectra. This spectro-temporal symmetry can be leveraged to selectively eliminate such contributions in action-detected 2DES and pump--probe spectroscopy in a system-independent way. Among these contributions belongs the infamous 'incoherent mixing', a stationary time-independent background of ground-state signals correlated by nonlinear excitation interaction during signal emission. As we have shown in detail, subtraction of positive- and negative-waiting-time spectra suppresses pulse-overlap and coherent signals, eliminates the incoherent mixing background, and highlights the excited state dynamics such as energy transfer. The approach can be applied to any existing action-detected spectroscopy experiment, without the need for experimental modifications. We believe that, due to its system-independent universality, the approach will find application in action-detected nonlinear spectroscopy in general and in F-PP and F-2DES in particular. Notably, it promises to open the door for fluorescence- and photocurrent-detected measurement of large coupled systems, such as photosynthetic complexes.

\section*{Supplementary material}
Supplementary material is available with the following sections. S1. Detailed derivation of the F-2DES expressions for normal and reversed pulse ordering. S2. explicit formulas for a coupled dimer. S3. Expressions for electronic coherence signals. S4. Diagrams for pulse overlap region. S5. Description of the numerical calculation of the F-PP signal of a model aggregate.   

\section*{Acknowledgements}

We thank Stephanie Sanders and Julian L\"uttig from the group of Jennifer Ogilvie for providing the F-2DES data of LH2, and for enlightening discussions of F-2DES. We also acknowledge Stefan M\"uller for useful discussions and sharing insights into F-PP data subtraction. We further thank Tom\'a\v s Man\v cal for help with the Quantarhei software. The authors acknowledge funding by Charles University (grant no. PRIMUS/24/SCI/007, to P.M.).

\section*{Author declarations}
\noindent Conflict of interest: The authors have no conflicts to disclose.

\medskip

\noindent Author contributions:

\noindent Kate\v{r}ina Charv\'atov\'a: Conceptualization, Data Curation, Formal Analysis, Methodology, Software, Visualization, Writing - Original Draft Preparation, Review \& Editing

\noindent Pavel Mal\'y: Conceptualization, Funding Acquisition, Methodology, Resources, Supervision, Writing - Original Draft Preparation, Review \& Editing

\section*{Data availability}

All experimental data have been  published in the cited works, for availability see therein. QuantaRhei is an open source software for calculations. Remaining data that support the findings of this study are available from the corresponding author upon reasonable request. 

\section*{References}
\bibliography{references}

\end{document}


\maketitle
\section{F-2DES signal derivations}

To compare the fluorescence-detected two-dimensional electronic spectroscopy (F-2DES) signal for time $T>0$ and $T<0$ between the pump and probe pulses, we begin with the most general expression based on perturbation theory. In the relevant contributions to the F-2DES signal, the sample interacts four times with the electric field, which consists of two probe and two pump pulses, each separated by different time delays. This ultimately leads to a population state. The final two interactions correspond to spontaneous emission of a photon, making this a sixth-order signal.\cite{sun2024interconnection} Nevertheless, thanks to the separation of timescales and the incoherent nature of spontanous emission, we can without loss of generality treat the F-2DES process as a form of four-wave mixing.\cite{maly2021fluorescence}

Fourth-order nonlinear contribution to the density matrix of a system perturbed by the interaction with the excitation pulses can be expressed as 
\begin{equation}
\begin{split}
    &\hat{\rho}^{(4)}(t')=\iiiint dt_4dt_3dt_2dt_1 R^{(4)}(t_4,t_3,t_2,t_1)E(t-t_4)\\
    &E(t' - t_4-t_3)E(t'-t_4-t_3-t_2)E(t'-t_4-t_3-t_2-t_1),
\end{split}
\label{rho4}
\end{equation}
\noindent where
\begin{equation}
    R^{(4)}(t_4,t_3,t_2,t_1)=Tr_B\{\check{U}_0(t_4)\check{\mathcal{V}}\check{U}_0(t_3)\check{\mathcal{V}}\check{U}_0(t_2)\check{\mathcal{V}}\check{U}_0(t_1)\check{\mathcal{V}} \hat{\rho}_0 \}
\label{R4}
\end{equation}
is a the response function. $\check{\mathcal{V}}$ is a superoperator that mediates the matter-field interaction as $\check{\mathcal{V}}\hat{\rho}=\frac{i}{\hbar}\left[\hat{\mu},\hat{\rho}\right]$, with $\hat{\mu}$ being the molecular transition dipole operator. $\check{U}_0$ is the evolution superoperator of the whole system, and $Tr_B$ denotes the trace over the bath degrees of freedom.
$t_1,t_2,t_3,t_4$ are times between interactions with electric field $E(t)$ of the pulses. In time $t'$, the fluorescence signal is emitted with rate $K_{Fn}$.

\begin{equation}
    \textrm{FL}(t')=Tr\left(\sum_{n} |n\rangle K_{Fn} \langle n| \rho^{(4)}(t') \right)
\label{Fluorescence_signal_calculation}
\end{equation}
For our four pulse experiment, the electric field could be written as 
\begin{align}
     E(t')&=\mathcal{E}^{(+)}_{Pu_1}(t'+T+\tau+t)e^{i\varphi_{1}-i\omega_{Pu_1}(t'+T+\tau+t)}\\ \notag
     &+\mathcal{E}^{(+)}_{Pu_2}(t'+T+t)e^{i\varphi_{2}-i\omega_{Pu_2}(t'+T+t)}\\ \notag
     &+\mathcal{E}^{(+)}_{Pr_1}(t'+t)e^{i\varphi_{3}-i\omega_{Pr_1}(t'+t)}\\ \notag
     &+\mathcal{E}^{(+)}_{Pr_2}(t')e^{i\varphi_{4}-i\omega_{Pr_2}(t')} + c.c.,  
 \label{Epulseswhole}
\end{align}
where $\mathcal{E}_i(t')$ is the envelope of the $i$ -th pulse, $\omega_i$ its mean frequency, and $\varphi_i$ its phase. 

Scanning over $\tau$ and $t$ gives us after Fourier transformation the frequency resolution with variables $\omega_{\tau}$ and $\omega_{t}$, respectively. Time resolution is obtained by scanning the time delay $T$ between the pump and the probe.
\begin{equation}
\begin{split}
    &\textrm{F-2DES}^{(4)}(\omega_t,T,\omega_{\tau})=\iint\limits_{-\infty}^{\infty}\int\limits_0^{\infty}\cdot\cdot\cdot\int\limits_0^{\infty} R^{(4)}(t_4,t_3,t_2,t_1)E(t'-t_4) E(t' - t_4-t_3) \times \\
    &\times E(t'-t_4-t_3-t_2)E(t'-t_4-t_3-t_2-t_1) e^{i\omega_tt} e^{i\omega_{\tau}\tau} dt_1 dt_2 dt_3 dt_4 dt' dt d\tau.
\end{split}
\label{FL4_s_t4}
\end{equation}
Now, we will derive the expression for F-2DES for the normal and reversed order of the pulses. To disentangle the pulse spectra and system response, we assume time-ordered pulses shorter than the relevant system dynamics in the waiting time $t_2$. 
Because relaxation to the ground state is slow compared to timescale connected with FWHM of the probe pulse, fluorescence signal obtained in time between interaction with the last probe and time of arrival of its center is negligible. Thus, we can extract evolution over $t_4$ from the response function. 
\begin{equation}
\begin{split}
    &R^{(4)}(t_4,t_3,t_2,t_1)\approx\\
    &\approx\sum_{ij}K_{Fi} U_{ij}(t_4)Tr\{|j\rangle\langle j|[\hat{\mu},\check{U}(t_3)[\hat{\mu},\check{U}(t_2)[\hat{\mu},\check{U}(t_1),[\hat{\mu},\hat{\rho}_0]]]] \}=\\
    &=\sum_{ij}K_{Fi} U_{ij}(t_4)R_j^{(3)}(t_3,t_2,t_1)
\end{split}
\label{R4approximationR3}
\end{equation}
Further, we will assume that the waiting time $T$ between the pump and probe is much longer than the pulse length, so that the pulse overlap can be neglected and evolution over time $t_2$ can be approximated over $T$. We would also suppose that the two pump pulses have the same spectrum envelope $\mathcal{E}_{Pu}(\omega)$ centered at $\omega_{Pu}$, and the same holds for probe pulses with envelope $\mathcal{E}_{Pr}(\omega)$ centered at $\omega_{Pr}$.

\begin{figure}
\centering
\includegraphics[width=0.55\textwidth]{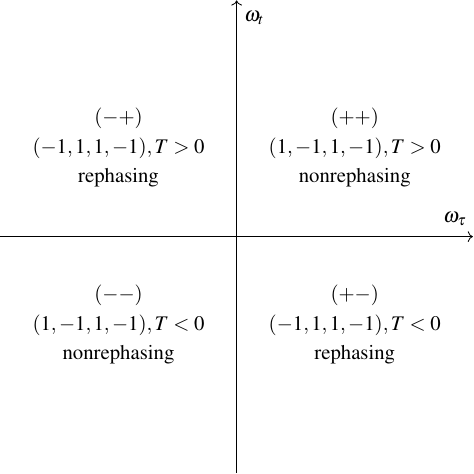}
\caption{Diagram for spectral position of the nonrephasing (1,-1,1,-1) and rephasing (-1,1,1,-1) signals in the F-2DES$(\omega_t,T>0,\omega_{\tau})$ map. Inverted time ordering of the pulses leads to the mirroring of the frequency quadrant.For opposite order of the pulses, the quadrants change. For example, the rephasing signal lies in the $(++)$ quadrant for $T>0$ but at $T<0$ appears in the $(- -)$ quadrant.}
\label{R_NR_diagram}
\end{figure}

We can distinguish four different cases - normal and opposite order of the pulses and for each rephasing and nonrephasing signal. Rephasing and nonrephasing case will be derived simultaneously using general signature $(\eta_1,\eta_2,\eta_3,\eta_4)$ as $(-1,1,1,-1)$ and $(1,-1,1,-1)$ respectively. 
Following the derivation introduced by Do et al.\cite{do2017simplified} for 2DES, we can write for standard time ordering of the pulses 
\begin{equation}
\begin{split}
    &\textrm{F-2DES}(\omega_t,T,\omega_{\tau})=\iint\limits_{-\infty}^{\infty}\int\limits_0^{\infty}\cdot\cdot\cdot\int\limits_0^{\infty} \sum_{ij}K_{Fi} U_{ij}(t_4)R_j^{(3)}(t_3,T,t_1)\times\\
    &\times \mathcal{E}^{(\eta_4)}_{Pr}(t'-t_4) e^{-i\eta_4\omega_{Pr}(t'-t_4)}\times\\
    &\times \mathcal{E}^{(\eta_3)}_{Pr}(t'+t-t_4-t_3) e^{-i\eta_3\omega_{Pr}(t'+t-t_4-t_3)}\times\\
    &\times \mathcal{E}^{(\eta_2)}_{Pu}(t'+t+T-t_4-t_3-t_2) e^{-i\eta_2\omega_{Pu}(t'+t+T-t_4-t_3-t_2)}\times\\
    &\times \mathcal{E}^{(\eta_1)}_{Pu}(t'+t+T+\tau-t_4-t_3-t_2-t_1) e^{-i\eta_1\omega_{Pu}(t'+t+T+\tau-t_4-t_3-t_2-t_1)}\times\\
    &\times e^{i\omega_tt} e^{i\omega_{\tau}\tau} dt_1 dt_2 dt_3 dt_4 dt' dt d\tau
\end{split}
\label{FL4_bez_t4t_2}
\end{equation}
To have the same variables as Ref. \cite{do2017simplified} we now transform our integrals as such $t_1=\tau_2-\tau_1$,$t_2=\tau_3-\tau_2$,$t_3=\tau_4-\tau_3$,$t_4=t'-\tau_4$ and also add corresponding $\Theta$-functions to ensure the correct order of the pulses, which further allows us to extend the limits of integration to from minus infinity to infinity.
\begin{equation}
\begin{split}
    &\textrm{F-2DES}(\omega_t,T,\omega_{\tau})=\int\limits_{0}^{\infty}\int\limits_{-\infty}^{\infty}\cdot\cdot\cdot\int\limits_{-\infty}^{\infty}\sum_{ij}K_{Fi} U_{ij}(t'-\tau_4)R_j^{(3)}(\tau_4-\tau_3,T,\tau_2-\tau_1)\times\\
    &\times \mathcal{E}^{(\eta_4)}_{Pr}(\tau_4) e^{-i\eta_4\omega_{Pr}\tau_4}\Theta(t'-\tau_4)\times\\
    &\times \mathcal{E}^{(\eta_3)}_{Pr}(t+\tau_3) e^{-i\eta_3\omega_{Pr}(t+\tau_3)}\Theta(\tau_4-\tau_3)\times\\
    &\times \mathcal{E}^{(\eta_2)}_{Pu}(t+T+\tau_2) e^{-i\eta_2\omega_{Pu}(t+T+\tau_2)}\times\\
    &\times \mathcal{E}^{(\eta_1)}_{Pu}(t+T+\tau+\tau_1) e^{-i\eta_1\omega_{Pu}(t+T+\tau+\tau_1)}\Theta(\tau_2-\tau_1)\times\\
    &\times e^{i\omega_tt} e^{i\omega_{\tau}\tau} d\tau_1 d\tau_2 d\tau_3 d\tau_4 dt' dt d\tau
\end{split}
\label{FL4_theta_fce}
\end{equation}
The only terms that depend on $\tau$ are the last pulse and factor for Fourier transform. Switching the order of integrations, we can write
\begin{equation}
\begin{split}
&\textrm{F-2DES}(\omega_t,T,\omega_{\tau})=\mathcal{E}^{(\eta_1)}_{Pu}(\eta_1(\omega_\tau-\eta_1\omega_{Pu}))\times\\
&\times\int\limits_{0}^{\infty}\int\limits_{-\infty}^{\infty}dt'd\tau_4 \sum_{ij}K_{Fi} U_{ij}(t'-\tau_4)\mathcal{E}^{(\eta_4)}_{Pr}(\tau_4) e^{-i\eta_4\omega_{Pr}\tau_4}\Theta(t'-\tau_4)\times\\
&\times \iint\limits_{-\infty}^{\infty}   dt  d\tau_3  \mathcal{E}^{(\eta_3)}_{Pr}(t+\tau_3) e^{-i\eta_3\omega_{Pr}(t+\tau_3)} e^{i\omega_tt}\Theta(\tau_4-\tau_3) \times\\
&\times \int\limits_{-\infty}^{\infty} d\tau_2 \mathcal{E}^{(\eta_2)}_{Pu}(t+T+\tau_2) e^{-i\eta_2\omega_{Pu}(t+T+\tau_2)}\times\\
&\times\int\limits_{-\infty}^{\infty} d\tau_1 R_j^{(3)}(\tau_4-\tau_3,T,\tau_2-\tau_1) \Theta(\tau_2-\tau_1) e^{-i\omega_{\tau}(t+T+\tau_1)}
\end{split}
\label{FL4_FT_tau}
\end{equation}
Further we proceed with integration over $\tau_1$
\begin{equation}
\begin{split}
    &\textrm{F-2DES}(\omega_t,T,\omega_{\tau})=\mathcal{E}^{(\eta_1)}_{Pu}(\eta_1(\omega_\tau-\eta_1\omega_{Pu}))\times\\
    &\times\int\limits_{0}^{\infty}\int\limits_{-\infty}^{\infty}dt'd\tau_4 \sum_{ij}K_{Fi} U_{ij}(t'-\tau_4)\mathcal{E}^{(\eta_4)}_{Pr}(\tau_4) e^{-i\eta_4\omega_{Pr}(\tau_4)}\Theta(t'-\tau_4)\times\\
    &\times \iint\limits_{-\infty}^{\infty}   dt  d\tau_3  \mathcal{E}^{(\eta_3)}_{Pr}(t+\tau_3) e^{-i\eta_3\omega_{Pr}(t+\tau_3)} e^{i\omega_tt}R_j^{(3)}(\tau_4-\tau_3,T,\omega_\tau)\Theta(\tau_4-\tau_3) \times\\
    &\times \int\limits_{-\infty}^{\infty} d\tau_2 \mathcal{E}^{(\eta_2)}_{Pu}(t+T+\tau_2) e^{-i\eta_2\omega_{Pu}(t+T+\tau_2)} e^{-i\omega_{\tau}(t+T+\tau_2)}
\end{split}
\label{FL4_int_tau1}
\end{equation}
Integration over $\tau_2$ leads to
\begin{equation}
\begin{split}
    &\textrm{F-2DES}(\omega_t,T,\omega_{\tau})=\mathcal{E}^{(\eta_1)}_{Pu}(\eta_1(\omega_\tau-\eta_1\omega_{Pu})) \mathcal{E}^{(\eta_2)}_{Pu}(-\eta_2(\eta_2\omega_{Pu}+\omega_\tau)) \times \\
    &\times \int\limits_{0}^{\infty}\int\limits_{-\infty}^{\infty}dt'd\tau_4 \sum_{ij}K_{Fi} U_{ij}(t'-\tau_4)\mathcal{E}^{(\eta_4)}_{Pr}(\tau_4) e^{-i\eta_4\omega_{Pr}(\tau_4)}\Theta(t'-\tau_4)\times\\
    &\times \iint\limits_{-\infty}^{\infty}   dt  d\tau_3  \mathcal{E}^{(\eta_3)}_{Pr}(t+\tau_3) e^{-i\eta_3\omega_{Pr}(t+\tau_3)} e^{i\omega_tt}R_j^{(3)}(\tau_4-\tau_3,T,\omega_\tau)\Theta(\tau_4-\tau_3)
\end{split}
\label{FL4_int_tau2}
\end{equation}
After Fourier transform over $t$ we get 
\begin{equation}
\begin{split}
    &\textrm{F-2DES}(\omega_t,T,\omega_{\tau})=\mathcal{E}^{(\eta_1)}_{Pu}(\eta_1(\omega_\tau-\eta_1\omega_{Pu})) \mathcal{E}^{(\eta_2)}_{Pu}(-\eta_2(\eta_2\omega_{Pu}+\omega_\tau))  \mathcal{E}^{(\eta_3)}_{Pr}(\eta_3(\omega_t-\eta_3\omega_{Pr}))\times \\
    &\times \int\limits_{0}^{\infty}\int\limits_{-\infty}^{\infty}dt'd\tau_4 \sum_{ij}K_{Fi} U_{ij}(t'-\tau_4)\mathcal{E}^{(\eta_4)}_{Pr}(\tau_4) e^{-i\eta_4\omega_{Pr}(\tau_4)}\Theta(t'-\tau_4)\times\\
    &\times \int\limits_{-\infty}^{\infty} d\tau_3 R_j^{(3)}(\tau_4-\tau_3,T,\omega_\tau)\Theta(\tau_4-\tau_3)  e^{-i\omega_t\tau_3} 
\end{split}
\label{FL4_int_t}
\end{equation}
After integration over $\tau_3$ further gives us
\begin{equation}
\begin{split}
    &\textrm{F-2DES}(\omega_t,T,\omega_{\tau})=\mathcal{E}^{(\eta_1)}_{Pu}(\eta_1(\omega_\tau-\eta_1\omega_{Pu})) \mathcal{E}^{(\eta_2)}_{Pu}(-\eta_2(\eta_2\omega_{Pu}+\omega_\tau))  \mathcal{E}^{(\eta_3)}_{Pr}(\eta_3(\omega_t-\eta_3\omega_{Pr}))\times \\
    &\times \sum_{ij}K_{Fi} R_j^{(3)}(\omega_t,T,\omega_\tau) \int\limits_{0}^{\infty}\int\limits_{-\infty}^{\infty}dt'd\tau_4  U_{ij}(t'-\tau_4) \mathcal{E}^{(\eta_4)}_{Pr}(\tau_4) e^{-i\eta_4\omega_{Pr}(\tau_4)}e^{-i\omega_t\tau_4} \Theta(t'-\tau_4)
\end{split}
\label{FL4_int_tau3}
\end{equation}
Finally, we approximate $U_{ij}(t'-\tau_4) \approx U_{ij}(t')$ and omit the Heaviside function, which is possible, because evolution $U_{ij}(t'-\tau_4)$ is slow compared to time scale connected to FWHM of used pulses. And consequently, we integrate over $\tau_4$ and $t'$:
\begin{equation}
\begin{split}
    &\textrm{F-2DES}(\omega_t,T>0,\omega_{\tau})=\sum_{j}\Phi_j R_j^{(3)}(\omega_t,T,\omega_\tau) \times\\
    &\times \mathcal{E}^{(\eta_1)}_{Pu}(\eta_1(\omega_\tau-\eta_1\omega_{Pu})) \mathcal{E}^{(\eta_2)}_{Pu}(-\eta_2(\omega_\tau+\eta_2\omega_{Pu}))  \mathcal{E}^{(\eta_3)}_{Pr}(\eta_3(\omega_t-\eta_3\omega_{Pr})) \mathcal{E}^{(\eta_4)}_{Pr}(-\eta_4(\omega_t+\eta_4\omega_{Pr}))\times\\
    &\times e^{i(\eta_1\varphi_1+\eta_2\varphi_2+\eta_3\varphi_3+\eta_4\varphi_4)}
\end{split}
\label{FL4_int_tau4t'}
\end{equation}
For both rephasing and nonrephasing contributions holds that $\eta_2=-\eta_1$, $\eta_4=-\eta_3$ and $\eta_i^2=1 \; (\forall i)$. Further we can use $I_x(\omega)=\mathcal{E}_x^{(+)}(\omega)\mathcal{E}_x^{(-)}(\omega)$, where $x\in \{\textrm{Pu, Pr}\}$, which significantly simplifies the equations
\bigskip
\begin{equation}
\begin{split}
    &\textrm{F-2DES}(\omega_t,T>0,\omega_{\tau})=\sum_j \Phi_j R_j^{(3)}(\omega_t,T,\omega_\tau)\times\\
    &\times I_{Pu}(\eta_1\omega_\tau-\omega_{Pu}) I_{Pr}(\eta_3\omega_t-\omega_{Pr}) e^{i(\eta_1(\varphi_1-\varphi_2)+\eta_3(\varphi_3-\varphi_4))}
\end{split}
\label{F2DES_general_positive}
\end{equation}
Upon inverting the pulse order and repeating the derivation, we obtain
\begin{equation}
\begin{split}
    &\textrm{F-2DES}(\omega_t,T<0,\omega_{\tau})=\sum_j \Phi_j R_j^{(3)}(\omega_\tau,T,\omega_t) \times\\
    &\times I_{Pu}(-\eta_1\omega_\tau-\omega_{Pu})I_{Pr}(-\eta_3\omega_t-\omega_{Pr}) e^{i(\eta_1(\varphi_1-\varphi_2)+\eta_3(\varphi_3-\varphi_4))}
\end{split}
\label{F2DES_general_negative}
\end{equation}
Here, the equations hold for a general pathway with signature $(\eta_1,\eta_2,\eta_3,\eta_4)$, which can be (-1,1,1,-1) for rephasing and (1,-1,1,-1) for nonrephasing. 
Explicitly, for rephasing spectrum we get
\begin{equation}
\begin{split}
    &\textrm{F-2DES}_{\textrm{R}}(\omega_t,T>0,\omega_{\tau})=\sum_j \Phi_j R_{j_{R}}^{(3)}(\omega_t,T,-\omega_\tau)\times\\
    &\times I_{Pu}(\omega_\tau-\omega_{Pu}))  I_{Pr}(\omega_t-\omega_{Pr}) e^{i(-\varphi_1+\varphi_2+\varphi_3-\varphi_4)}
\end{split}
\label{F2DES_rephasing_constant_positive}
\end{equation}
and
\begin{equation}
\begin{split}
    &\textrm{F-2DES}_{\textrm{R}}(\omega_t,T<0,\omega_{\tau})=\sum_j \Phi_j R_{j_{R}}^{(3)}(\omega_\tau,T,-\omega_t)\times\\
    &\times  I_{Pu}(\omega_\tau-\omega_{Pu}) I_{Pr}(\omega_t-\omega_{Pr}) e^{i(-\varphi_1+\varphi_2+\varphi_3-\varphi_4)},
\end{split}
\label{F2DES_rephasing_constant_negative}
\end{equation}
The negative frequency sign is given by the rephasing character of the response, as known in 2DES.\cite{hamm2011concepts} For nonrephasing spectrum, we have 
\begin{equation}
\begin{split}
    &\textrm{F-2DES}_{\textrm{NR}}(\omega_t,T>0,\omega_{\tau})=\sum_j \Phi_j R_{j_{NR}}^{(3)}(\omega_t,T,\omega_\tau)\times\\
    &\times I_{Pu}(\omega_\tau-\omega_{Pu})  I_{Pr}(\omega_t-\omega_{Pr}) e^{i(\varphi_1-\varphi_2+\varphi_3-\varphi_4)}
\end{split}
\label{F2DES_nonrephasing_constant_positive}
\end{equation}
and
\begin{equation}
\begin{split}
    &\textrm{F-2DES}_{\textrm{NR}}(\omega_t,T<0,\omega_{\tau})=\sum_j \Phi_j R_{j_{NR}}^{(3)}(-\omega_\tau,T,-\omega_t)\times\\
    &\times  I_{Pu}(\omega_\tau-\omega_{Pu}) I_{Pr}(\omega_t-\omega_{Pr}) e^{i(\varphi_1-\varphi_2+\varphi_3-\varphi_4)}
\end{split}
\label{F2DES_nonrephasing_constant_negative}
\end{equation}
We can see that upon inverting the time ordering of the pulses, we get a signal where all signs of $\omega_{t}$ or $\omega_{\tau}$ are opposite than for $T>0$, but the phase signature remains the same. As known from standard 2DES, the rephasing and non-rephasing spectra end up in different sign quadrants in frequency.\cite{hamm2011concepts} Upon time reversal, the time and thus frequency axes are mirrored, as shown in Fig.\ref{R_NR_diagram}, thus every contribution in $T>0$ has its counterpart in $T<0$. This is largely a formality, when isolating the rephasing or nonrephasing spectra one has to choose the appropriate quadrant. For example, to get subtracted nonrephasing spectra with phase signature $(1,-1,1,-1)$, we have to take signal in quadrant $(++)$ for $T>0$ and subtract signal in $(--)$ for $T<0$. For directly measured absorptive spectra, the sign is automatically correct same as in 2DES in pump--probe geometry.\cite{fuller2015experimental}

\bigskip

For signals with faster dynamics in waiting time between pump  and probe, we have to proceed more carefully. Following Do et al. \cite{do2017simplified} further for consistency, we take the time evolution of the response in $t_2$ to be of the form $e^{-i\Xi_{ij} t_2}=e^{-i\Xi_{ij} (\tau_3-\tau_2)}$. Here $\Xi_{ij}=\Omega_{ij}+i\gamma_{ij}$, $\Omega$ standing for frequency of oscillation $ij$ and $\gamma_{ij}$ is its dephasing rate. Following the same procedure, we get for classical order of the pulses:
\begin{equation}
\begin{split}
    &\textrm{F-2DES}(\omega_t,T>0,\omega_{\tau})=\sum_{ij} \Phi_i R_{ij}^{(3)}(\omega_t,\omega_\tau)e^{-i\Xi_{ij} |T|}  \times\\
    &\times\mathcal{E}^{(\eta_1)}_{Pu}(-\eta_1\omega_\tau-\omega_{Pu}) \mathcal{E}^{(-\eta_1)}_{Pu}(-\eta_1(\omega_\tau-\Xi_{ij})-\omega_{Pu})\\  
  &\mathcal{E}^{(\eta_3)}_{Pr}(-\eta_3(\omega_t-\Xi_{ij})-\omega_{Pr}) \mathcal{E}^{(-\eta_3)}_{Pr}(-\eta_3\omega_t-\omega_{Pr}),
\end{split}
\label{F2DES_general_for_oscilations_positive}
\end{equation}
where $ R_{ij}^{(3)}(\omega_t,\omega_\tau)$ is response function with $|i\rangle \langle j|$ coherence in waiting time $|T|$ and ending in population $|i\rangle \langle i|$. 
For opposite time ordering of the pulses:
\begin{equation}
\begin{split}
    &\textrm{F-2DES}(\omega_t,T<0,\omega_{\tau})=\sum_{ij} \Phi_i R_{ij}^{(3)}(\omega_\tau,\omega_t)e^{-i\Xi_{ij} |T|} \\
    &\mathcal{E}^{(\eta_1)}_{Pu}(\eta_1\omega_\tau-\omega_{Pu})\mathcal{E}^{(-\eta_1)}_{Pu}(\eta_1(\omega_\tau-\Xi_{ij})-\omega_{Pu})  \times\\
    &\times \mathcal{E}^{(\eta_3)}_{Pr}(\eta_3(\omega_t-\Xi_{ij})-\omega_{Pr}) \mathcal{E}^{(-\eta_3)}_{Pr}(\eta_3\omega_t-\omega_{Pr}))
\end{split}
\label{F2DES_general_for_oscilations_negative}
\end{equation}
Comparing these two formulae, we can see that for pulses which are spectrally narrow enough that $\mathcal{E}^{(\eta_i)}_i(\omega-\Xi)\not\approx \mathcal{E}^{(\eta_i)}_i(\omega)$, oscillations are not symmetric around time zero and could not be subtracted.

\section{Excitonic system signal}
\subsection{General case}

Here, we present the equations for the F-2DES signal of a general excitonic system with two-excitation signal yield $\Gamma$ to point out some of its properties. 

\begin{figure}
\includegraphics[width=1.0\textwidth]{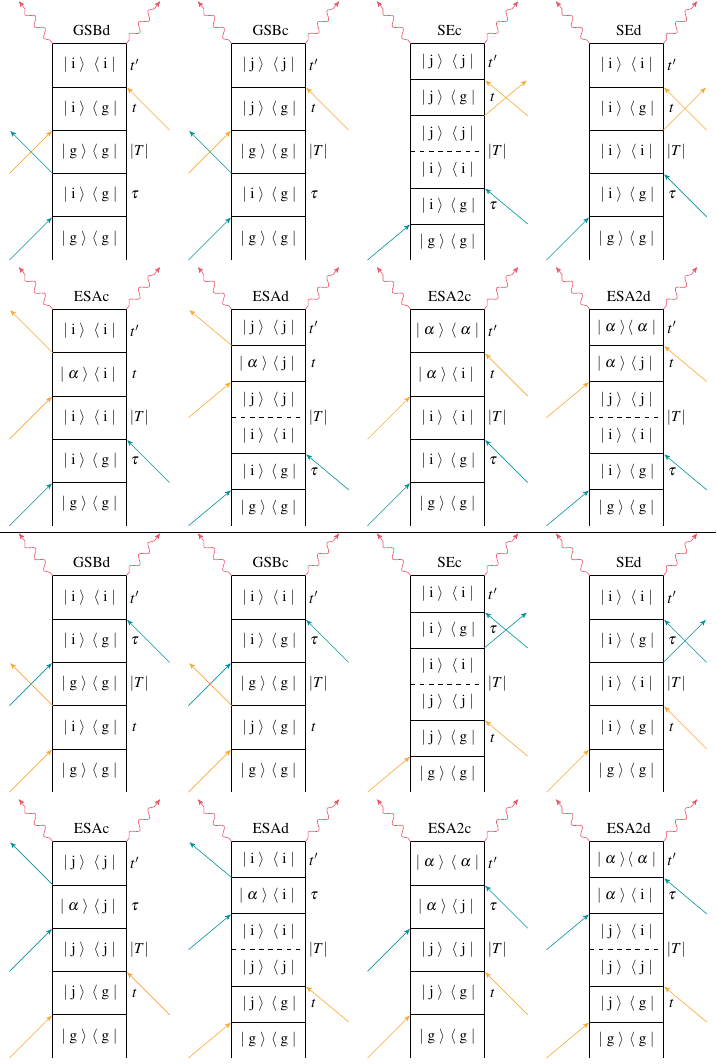}
\caption{Feynman diagrams of nonrephasing response pathway contributions of F-2DES with phase signature (1,-1,1,-1), for $T>0$ and $T<0$.}
\label{FD}
\end{figure}

All the relevant response pathways are shown in form of double-sided Feynman diagrams in Fig. (\ref{FD}). From these, we can construct the total response. Doing so, we consider factorization between the three time-delay intervals. The optical coherence $ig$ gives rise to a lineshape $G^g_{ig}(\omega)$ if the system was previously in its ground state, and to a lineshape $G^e_{ig}(\omega)$ if the system was in the excited state. The transition to a two-exciton state $\alpha$ leads to a lineshape $G^f_{\alpha i}(\omega)$. Each interaction with the field goes with the corresponding element of the transition dipole moment such as $\mu_{ig}$ or $\mu_{\alpha i}$. Finally, the time evolution in the middle interval is governed by the propagator $\mathcal{U} (T)$, that reflects population transfer from state $i$ to state $k$, $\mathcal{U}_{ki} (T) = \mathcal{U}_{kkii} (T)$, as well as coherent oscillations. The sign of the diagram corresponds to the number of interactions from the right, due to the corresponding commutator. From these building blocks, we can construct the signal for $T>0$:
\begin{equation}
\begin{split}
&\textrm{F-2DES}(\omega_{{\tau}},T>0,\omega_{{t}})=\\
&-\sum_{ij}|\mu_i|^2|\mu_j|^2G^g_{jg}(\omega_{t})G^g_{ig}(\omega_{\tau})\\
&-\sum_{ij}|\mu_i|^2|\mu_j|^2\mathcal{U}_{ji}(|T|)G^e_{jg}(\omega_{t})G^g_{ig}(\omega_{\tau})\\
&+(\Gamma-1)\sum_{ik\alpha}|\mu_{\alpha k}|^2|\mu_j|^2\mathcal{U}_{ki}(|T|)G^f_{\alpha,k}(\omega_{t})G^g_{ig}(\omega_{\tau})\\
&+R^{(4)}_\textrm{coh}(0,\omega_{t},T,\omega_{\tau})\\
\label{F-2DES_explicit_positive}
\end{split}
\end{equation}
For $\Gamma=1$ the last four contributions (ESA1 and ESA2) cancel.
For negative times, the signal is as follows
\begin{equation}
\begin{split}
&\textrm{F-2DES}(\omega_{{\tau}},T<0,\omega_{{t}})=\\
&-\sum_{ij}|\mu_i|^2|\mu_j|^2G^g_{ig}(\omega_{\tau})G^g_{jg}(\omega_t)\\
&-\sum_{ij}|\mu_i|^2|\mu_j|^2\mathcal{U}_{ij}(|T|)G^e_{ig}(\omega_\tau)G^g_{jg}(\omega_t)\\
&+(\Gamma-1)\sum_{jk\alpha}|\mu_{\alpha k}|^2|\mu_j|^2\mathcal{U}_{kj}(|T|)G^f_{\alpha,k}(\omega_\tau)G^g_{jg}(\omega_t)\\
&+R^{(4)}_\textrm{coh}(0,\omega_{\tau},T,\omega_{t})\\
\label{F-2DES_explicit_negative}
\end{split}
\end{equation}
Again, for $\Gamma=1$ ESA1 and ESA2 contributions cancel each other. Thanks to the stationary signal from the ground state, GSB contributions are symmetric around $T=0$ and can be subtracted. 

The subtracted signal is thus (for $\Gamma=1$)
\begin{equation}
\begin{split}
&\textrm{F-2DES}_{\textrm{diff}}(\omega_{{\tau}},|T|,\omega_{{t}})=\\
&\sum_{ij}|\mu_i|^2|\mu_j|^2 \left(\mathcal{U}_{ij}(|T|)G_{jg}^{g}(\omega_{t})G_{ig}^{e}(\omega_{\tau})-\mathcal{U}_{ji}(|T|)G_{jg}^{e}(\omega_{t})G_{ig}^{g}(\omega_{\tau})\right)\\
&+\left(R^{(4)}_\textrm{coh}(0,\omega_{t},|T|,\omega_{\tau})-R^{(4)}_\textrm{coh}(0,\omega_{\tau},|T|,\omega_{t})\right)
\label{F-2DES_explicit_diff}
\end{split}
\end{equation}

\subsection{Coupled dimer}

The F-2DES spectra are schematically shown in Fig. 3 in the main text for a heterodimer with downhill energy transfer, where $K_{12}>>K_{21}$. In this case, we can neglect the energy transfer from $1$ to $2$ and assume only that from $2$ to $1$. Then we can approximate the evolution propagator as $\mathcal{U}_{12}(|T|)=1-e^{-K_{12}|T|}$ and to $\mathcal{U}_{22}(|T|)=e^{-K_{12}|T|}$. Accordingly, $\mathcal{U}_{21}(|T|)=0$ and $\mathcal{U}_{11}(|T|)=1$. 
F-2DES spectra in positive waiting times thus simplify to 
\begin{equation}
\begin{split}
\textrm{F-2DES}(\omega_{{\tau}},T>0,\omega_{{t}})=&\underbrace{-\sum^{2}_{i,j=1}|\mu_i|^2|\mu_j|^2G^g_{jg}(\omega_{t})G^g_{ig}(\omega_{\tau})}_{\textrm{GSBc}+\textrm{GSBd}}\\
&\underbrace{-|\mu_1|^2|\mu_2|^2(1-e^{-K_{12}|T|})G^e_{1g}(\omega_{t})G^g_{2g}(\omega_{\tau})}_{\textrm{SEc}}\\
&\underbrace{-|\mu_2|^4e^{-K_{12}|T|}G^e_{2g}(\omega_{t})G^g_{2g}(\omega_{\tau})-|\mu_1|^4G^e_{1g}(\omega_{t})G^g_{1g}(\omega_{\tau})}_{\textrm{SEd}}\\
&\underbrace{(\Gamma-1)|\mu_2|^4(1-e^{-K_{12}|T|})G^e_{12,1}(\omega_{t})G^g_{2g}(\omega_{\tau})}_{\textrm{ESAd}}\\
&\underbrace{(\Gamma-1)\left(|\mu_1|^2|\mu_2|^2e^{-K_{12}|T|}G^e_{21,2}(\omega_{t})G^g_{2g}(\omega_{\tau})+|\mu_1|^2|\mu_2|^2G^e_{12,1}(\omega_{t})G^g_{1g}(\omega_{\tau})\right)}_{\textrm{ESAc}}\\
&\underbrace{+R^{(4)}_\textrm{coh}(0,\omega_{t},|T|,\omega_{\tau})}_{\textrm{oscillations}}\\
\label{F-2DES_dimer_positive}
\end{split}
\end{equation}
we can see that the signal of peak $G^e_{2g}(\omega_{t})G^g_{2g}(\omega_{\tau})$ is decreasing in time on behalf of the rise of off-diagonal peak $G^e_{1g}(\omega_{t})G^g_{2g}(\omega_{\tau})$. In this way, we can observe energy transfer in F-2DES. 
For F-2DES in $T<0$ we have in this approximation
\begin{equation}
\begin{split}
&\textrm{F-2DES}(\omega_{{\tau}},T<0,\omega_{{t}})=\underbrace{-\sum^{2}_{i,j=1}|\mu_i|^2|\mu_j|^2G^g_{jg}(\omega_{\tau})G^g_{ig}(\omega_t)}_{\textrm{GSBc+GSBd}}\\
&\underbrace{-|\mu_1|^2|\mu_2|^2(1-e^{-K_{12}|T|})G^e_{1g}(\omega_\tau)G^g_{2g}(\omega_t)}_{\textrm{SEc}}\\
&\underbrace{-|\mu_2|^4e^{-K_{12}|T|}G^e_{2g}(\omega_\tau)G^g_{2g}(\omega_t)-|\mu_1|^4G^e_{1g}(\omega_\tau)G^g_{1g}(\omega_t)}_{\textrm{SEd}}\\
&\underbrace{+(\Gamma-1)|\mu_2|^4(1-e^{-K_{12}|T|})G^e_{12,1}(\omega_\tau)G^g_{2g}(\omega_t)}_{\textrm{ESAd}}\\
&\underbrace{+(\Gamma-1)\left(|\mu_1|^2|\mu_2|^2e^{-K_{12}|T|}G^e_{21,2}(\omega_\tau)G^g_{2g}(\omega_t)+|\mu_1|^2|\mu_2|^2G^e_{12,1}(\omega_\tau)G^g_{1g}(\omega_t)\right)}_{\textrm{ESAc}}\\
&\underbrace{+R^{(4)}_\textrm{coh}(0,\omega_{\tau},|T|,\omega_{t})}_{\textrm{oscillations}}\\
\label{F-2DES_dimer_negative}
\end{split}
\end{equation}
For the special case of equal oscillator strengths $\mu_1=\mu_2$, the dynamics of the F-PP signal, obtained by integration over $\omega_\tau$, at $T<0$ cancels and the signal is constant (on scales, where both radiative and nonradiative decay to the ground state is negligible, i.e. $e^{-K_{R}|T|}\approx 1$). In other cases, we can see the transfer between chromophores even for negative waiting time. This holds also if we, for example, spectrally select only some chromophores (which is effectively similar to changing their oscillator strength).

After subtraction of the negative-time data from the positive-time spectra, we obtain the difference signal 
\begin{equation}
\begin{split}
&\textrm{F-2DES}_{\textrm{diff}}(\omega_{{\tau}},|T|,\omega_{{t}})=\\
&|\mu_1|^2|\mu_2|^2\times\\
&\times \left(1-e^{-K_{12}|T|}\right)\left(G_{1g}^{e}(\omega_{t})G_{2g}^{g}(\omega_{\tau})-G_{2g}^{g}(\omega_{t})G_{1g}^{e}(\omega_{\tau})\right)+\\
&+|\mu_2|^2e^{-K_{12}|T|}\left(G_{2g}^{e}(\omega_{t})G_{2g}^{g}(\omega_{\tau})-G_{2g}^{g}(\omega_{t})G_{2g}^{e}(\omega_{\tau})\right)\\
&+|\mu_1|^2\left(G_{1g}^{e}(\omega_{t})G_{1g}^{g}(\omega_{\tau})-G_{1g}^{g}(\omega_{t})G_{1g}^{e}(\omega_{\tau})\right)\\
&+R^{(4)}_\textrm{coh}(0,\omega_{t},|T|,\omega_{\tau})-R^{(4)}_\textrm{coh}(0,\omega_{\tau},|T|,\omega_{t})
\label{F-2DES_dimer_diff}
\end{split}
\end{equation}
Ground-state bleach contributions canceled each other, thus only stimulated emission and oscillations contribute to the signal difference. When we integrate over area large enough, so it contains both absorption and emission lineshape of the corresponding peak, the diagonal peaks in the F-2DES signal contribute with zero traces and the oscillations get canceled from all traces, as shown in section \ref{sec:oscillations}. The energy transfer is observable in the dynamics of off-diagonal peaks, which are now without the unwanted constant background.

\section{Oscillations}
\label{sec:oscillations}
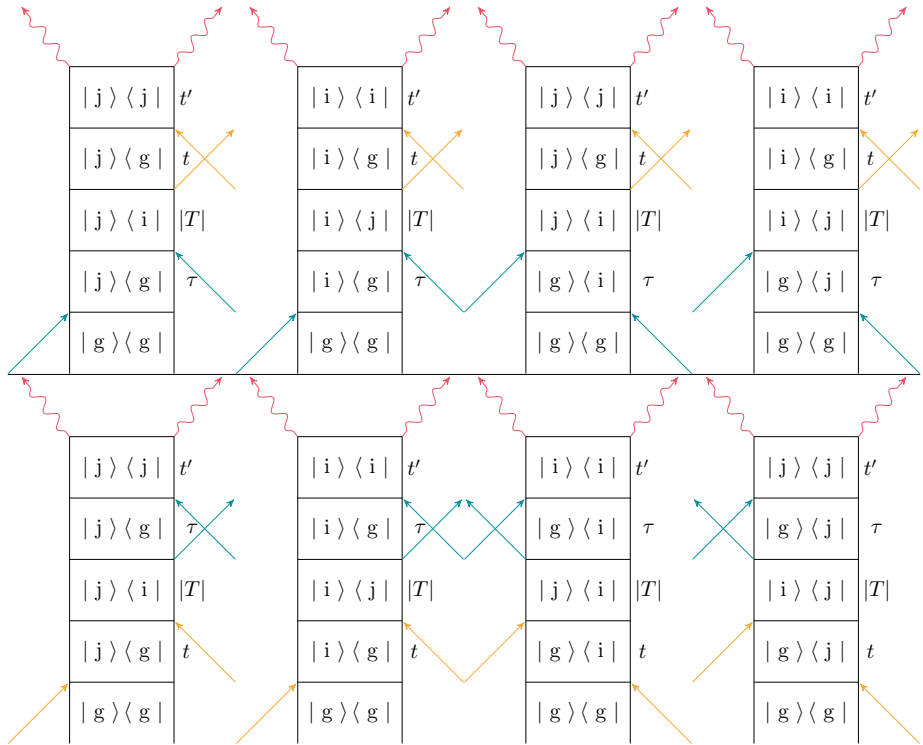
\begin{figure}

\setlength{\tabcolsep}{0mm} 
\def\arraystretch{0.0} 
\centering
\resizebox{\linewidth}{!}{%
\begin{tabular}{ c c c c}

\begin{tikzpicture}

\draw[] (0,0) -- (0,5) -- (1.7,5) -- (1.7,0);
\draw[] (0,1) -- (1.7,1);
\draw[] (0,2) -- (1.7,2);
\draw[] (0,3) -- (1.7,3);
\draw[] (0,4) -- (1.7,4);

\draw[->,> = stealth', shorten > = 1pt,mojemodra] (2.7,3) -- (1.7,4); 
\draw[->,> = stealth', shorten > = 1pt,mojemodra] (1.7,3) -- (2.7,4); 
\draw[->,> = stealth', shorten > = 1pt,mojezluta] (2.7,1) -- (1.7,2); 
\draw[->,> = stealth', shorten > = 1pt,mojezluta] (-1,0) -- (0,1); 

\draw[->,> = stealth', shorten > = 1pt,auto,decorate, decoration={snake, pre length=3pt,post length=3pt},mojecervena] (0,5) -- (-0.8,6);
\draw[->,> = stealth', shorten > = 1pt,auto,decorate, decoration={snake, pre length=3pt,post length=3pt},mojecervena] (1.7,5) -- (2.5,6);

\node at (0.5,4.5) {$|$ j $\rangle$};
\node at (0.5,3.5) {$|$ j $\rangle$};
\node at (0.5,2.5) {$|$ j $\rangle$};
\node at (0.5,1.5) {$|$ j $\rangle$};
\node at (0.5,0.5) {$|$ g $\rangle$};

\node at (1.2,4.5) {$\langle$ j $|$};
\node at (1.2,3.5) {$\langle$ g $|$};
\node at (1.2,2.5) {$\langle$ i $|$};
\node at (1.2,1.5) {$\langle$ g $|$};
\node at (1.2,0.5) {$\langle$ g $|$};

\node at (2.0,1.5) {$\tau$};
\node at (2.0,2.5) {$|T|$};
\node at (1.9,3.5) {$t$};
\node at (1.9,4.5) {$t'$};
\end{tikzpicture}
&\begin{tikzpicture}

\draw[] (0,0) -- (0,5) -- (1.7,5) -- (1.7,0);
\draw[] (0,1) -- (1.7,1);
\draw[] (0,2) -- (1.7,2);
\draw[] (0,3) -- (1.7,3);
\draw[] (0,4) -- (1.7,4);

\draw[->,> = stealth', shorten > = 1pt,mojemodra] (2.7,3) -- (1.7,4); 
\draw[->,> = stealth', shorten > = 1pt,mojemodra] (1.7,3) -- (2.7,4); 
\draw[->,> = stealth', shorten > = 1pt,mojezluta] (2.7,1) -- (1.7,2); 
\draw[->,> = stealth', shorten > = 1pt,mojezluta] (-1,0) -- (0,1); 

\draw[->,> = stealth', shorten > = 1pt,auto,decorate, decoration={snake, pre length=3pt,post length=3pt},mojecervena] (0,5) -- (-0.8,6);
\draw[->,> = stealth', shorten > = 1pt,auto,decorate, decoration={snake, pre length=3pt,post length=3pt},mojecervena] (1.7,5) -- (2.5,6);

\node at (0.5,4.5) {$|$ i $\rangle$};
\node at (0.5,3.5) {$|$ i $\rangle$};
\node at (0.5,2.5) {$|$ i $\rangle$};
\node at (0.5,1.5) {$|$ i $\rangle$};
\node at (0.5,0.5) {$|$ g $\rangle$};

\node at (1.2,4.5) {$\langle$ i $|$};
\node at (1.2,3.5) {$\langle$ g $|$};
\node at (1.2,2.5) {$\langle$ j $|$};
\node at (1.2,1.5) {$\langle$ g $|$};
\node at (1.2,0.5) {$\langle$ g $|$};

\node at (2.0,1.5) {$\tau$};
\node at (2.0,2.5) {$|T|$};
\node at (1.9,3.5) {$t$};
\node at (1.9,4.5) {$t'$};
\end{tikzpicture}
&\begin{tikzpicture}

\draw[] (0,0) -- (0,5) -- (1.7,5) -- (1.7,0);
\draw[] (0,1) -- (1.7,1);
\draw[] (0,2) -- (1.7,2);
\draw[] (0,3) -- (1.7,3);
\draw[] (0,4) -- (1.7,4);

\draw[->,> = stealth', shorten > = 1pt,mojemodra] (2.7,3) -- (1.7,4); 
\draw[->,> = stealth', shorten > = 1pt,mojemodra] (1.7,3) -- (2.7,4); 
\draw[->,> = stealth', shorten > = 1pt,mojezluta] (2.7,0) -- (1.7,1); 
\draw[->,> = stealth', shorten > = 1pt,mojezluta] (-1,1) -- (0,2); 

\draw[->,> = stealth', shorten > = 1pt,auto,decorate, decoration={snake, pre length=3pt,post length=3pt},mojecervena] (0,5) -- (-0.8,6);
\draw[->,> = stealth', shorten > = 1pt,auto,decorate, decoration={snake, pre length=3pt,post length=3pt},mojecervena] (1.7,5) -- (2.5,6);

\node at (0.5,4.5) {$|$ j $\rangle$};
\node at (0.5,3.5) {$|$ j $\rangle$};
\node at (0.5,2.5) {$|$ j $\rangle$};
\node at (0.5,1.5) {$|$ g $\rangle$};
\node at (0.5,0.5) {$|$ g $\rangle$};

\node at (1.2,4.5) {$\langle$ j $|$};
\node at (1.2,3.5) {$\langle$ g $|$};
\node at (1.2,2.5) {$\langle$ i $|$};
\node at (1.2,1.5) {$\langle$ i $|$};
\node at (1.2,0.5) {$\langle$ g $|$};

\node at (2.0,1.5) {$\tau$};
\node at (2.0,2.5) {$|T|$};
\node at (1.9,3.5) {$t$};
\node at (1.9,4.5) {$t'$};
\end{tikzpicture}
&\begin{tikzpicture}

\draw[] (0,0) -- (0,5) -- (1.7,5) -- (1.7,0);
\draw[] (0,1) -- (1.7,1);
\draw[] (0,2) -- (1.7,2);
\draw[] (0,3) -- (1.7,3);
\draw[] (0,4) -- (1.7,4);

\draw[->,> = stealth', shorten > = 1pt,mojemodra] (2.7,3) -- (1.7,4); 
\draw[->,> = stealth', shorten > = 1pt,mojemodra] (1.7,3) -- (2.7,4); 
\draw[->,> = stealth', shorten > = 1pt,mojezluta] (2.7,0) -- (1.7,1); 
\draw[->,> = stealth', shorten > = 1pt,mojezluta] (-1,1) -- (0,2); 

\draw[->,> = stealth', shorten > = 1pt,auto,decorate, decoration={snake, pre length=3pt,post length=3pt},mojecervena] (0,5) -- (-0.8,6);
\draw[->,> = stealth', shorten > = 1pt,auto,decorate, decoration={snake, pre length=3pt,post length=3pt},mojecervena] (1.7,5) -- (2.5,6);

\node at (0.5,4.5) {$|$ i $\rangle$};
\node at (0.5,3.5) {$|$ i $\rangle$};
\node at (0.5,2.5) {$|$ i $\rangle$};
\node at (0.5,1.5) {$|$ g $\rangle$};
\node at (0.5,0.5) {$|$ g $\rangle$};

\node at (1.2,4.5) {$\langle$ i $|$};
\node at (1.2,3.5) {$\langle$ g $|$};
\node at (1.2,2.5) {$\langle$ j $|$};
\node at (1.2,1.5) {$\langle$ j $|$};
\node at (1.2,0.5) {$\langle$ g $|$};

\node at (2.0,1.5) {$\tau$};
\node at (2.0,2.5) {$|T|$};
\node at (1.9,3.5) {$t$};
\node at (1.9,4.5) {$t'$};
\end{tikzpicture}\\

\hline

\begin{tikzpicture}

\draw[] (0,0) -- (0,5) -- (1.7,5) -- (1.7,0);
\draw[] (0,1) -- (1.7,1);
\draw[] (0,2) -- (1.7,2);
\draw[] (0,3) -- (1.7,3);
\draw[] (0,4) -- (1.7,4);

\draw[->,> = stealth', shorten > = 1pt,mojezluta] (2.7,3) -- (1.7,4); 
\draw[->,> = stealth', shorten > = 1pt,mojezluta] (1.7,3) -- (2.7,4); 
\draw[->,> = stealth', shorten > = 1pt,mojemodra] (2.7,1) -- (1.7,2); 
\draw[->,> = stealth', shorten > = 1pt,mojemodra] (-1,0) -- (0,1); 

\draw[->,> = stealth', shorten > = 1pt,auto,decorate, decoration={snake, pre length=3pt,post length=3pt},mojecervena] (0,5) -- (-0.8,6);
\draw[->,> = stealth', shorten > = 1pt,auto,decorate, decoration={snake, pre length=3pt,post length=3pt},mojecervena] (1.7,5) -- (2.5,6);

\node at (0.5,4.5) {$|$ j $\rangle$};
\node at (0.5,3.5) {$|$ j $\rangle$};
\node at (0.5,2.5) {$|$ j $\rangle$};
\node at (0.5,1.5) {$|$ j $\rangle$};
\node at (0.5,0.5) {$|$ g $\rangle$};

\node at (1.2,4.5) {$\langle$ j $|$};
\node at (1.2,3.5) {$\langle$ g $|$};
\node at (1.2,2.5) {$\langle$ i $|$};
\node at (1.2,1.5) {$\langle$ g $|$};
\node at (1.2,0.5) {$\langle$ g $|$};

\node at (2.0,3.5) {$\tau$};
\node at (2.0,2.5) {$|T|$};
\node at (1.9,1.5) {$t$};
\node at (1.9,4.5) {$t'$};
\end{tikzpicture}
&\begin{tikzpicture}

\draw[] (0,0) -- (0,5) -- (1.7,5) -- (1.7,0);
\draw[] (0,1) -- (1.7,1);
\draw[] (0,2) -- (1.7,2);
\draw[] (0,3) -- (1.7,3);
\draw[] (0,4) -- (1.7,4);

\draw[->,> = stealth', shorten > = 1pt,mojezluta] (2.7,3) -- (1.7,4); 
\draw[->,> = stealth', shorten > = 1pt,mojezluta] (1.7,3) -- (2.7,4); 
\draw[->,> = stealth', shorten > = 1pt,mojemodra] (2.7,1) -- (1.7,2); 
\draw[->,> = stealth', shorten > = 1pt,mojemodra] (-1,0) -- (0,1); 

\draw[->,> = stealth', shorten > = 1pt,auto,decorate, decoration={snake, pre length=3pt,post length=3pt},mojecervena] (0,5) -- (-0.8,6);
\draw[->,> = stealth', shorten > = 1pt,auto,decorate, decoration={snake, pre length=3pt,post length=3pt},mojecervena] (1.7,5) -- (2.5,6);

\node at (0.5,4.5) {$|$ i $\rangle$};
\node at (0.5,3.5) {$|$ i $\rangle$};
\node at (0.5,2.5) {$|$ i $\rangle$};
\node at (0.5,1.5) {$|$ i $\rangle$};
\node at (0.5,0.5) {$|$ g $\rangle$};

\node at (1.2,4.5) {$\langle$ i $|$};
\node at (1.2,3.5) {$\langle$ g $|$};
\node at (1.2,2.5) {$\langle$ j $|$};
\node at (1.2,1.5) {$\langle$ g $|$};
\node at (1.2,0.5) {$\langle$ g $|$};

\node at (2.0,3.5) {$\tau$};
\node at (2.0,2.5) {$|T|$};
\node at (1.9,1.5) {$t$};
\node at (1.9,4.5) {$t'$};
\end{tikzpicture}

&\begin{tikzpicture}

\draw[] (0,0) -- (0,5) -- (1.7,5) -- (1.7,0);
\draw[] (0,1) -- (1.7,1);
\draw[] (0,2) -- (1.7,2);
\draw[] (0,3) -- (1.7,3);
\draw[] (0,4) -- (1.7,4);

\draw[->,> = stealth', shorten > = 1pt,mojezluta] (-1,3) -- (0,4); 
\draw[->,> = stealth', shorten > = 1pt,mojezluta] (0,3) -- (-1,4); 
\draw[->,> = stealth', shorten > = 1pt,mojemodra] (2.7,0) -- (1.7,1); 
\draw[->,> = stealth', shorten > = 1pt,mojemodra] (-1,1) -- (0,2); 

\draw[->,> = stealth', shorten > = 1pt,auto,decorate, decoration={snake, pre length=3pt,post length=3pt},mojecervena] (0,5) -- (-0.8,6);
\draw[->,> = stealth', shorten > = 1pt,auto,decorate, decoration={snake, pre length=3pt,post length=3pt},mojecervena] (1.7,5) -- (2.5,6);

\node at (0.5,4.5) {$|$ i $\rangle$};
\node at (0.5,3.5) {$|$ g $\rangle$};
\node at (0.5,2.5) {$|$ j $\rangle$};
\node at (0.5,1.5) {$|$ g $\rangle$};
\node at (0.5,0.5) {$|$ g $\rangle$};

\node at (1.2,4.5) {$\langle$ i $|$};
\node at (1.2,3.5) {$\langle$ i $|$};
\node at (1.2,2.5) {$\langle$ i $|$};
\node at (1.2,1.5) {$\langle$ i $|$};
\node at (1.2,0.5) {$\langle$ g $|$};

\node at (2.0,3.5) {$\tau$};
\node at (2.0,2.5) {$|T|$};
\node at (1.9,1.5) {$t$};
\node at (1.9,4.5) {$t'$};
\end{tikzpicture}
&\begin{tikzpicture}

\draw[] (0,0) -- (0,5) -- (1.7,5) -- (1.7,0);
\draw[] (0,1) -- (1.7,1);
\draw[] (0,2) -- (1.7,2);
\draw[] (0,3) -- (1.7,3);
\draw[] (0,4) -- (1.7,4);

\draw[->,> = stealth', shorten > = 1pt,mojezluta] (-1,3) -- (0,4); 
\draw[->,> = stealth', shorten > = 1pt,mojezluta] (0,3) -- (-1,4); 
\draw[->,> = stealth', shorten > = 1pt,mojemodra] (2.7,0) -- (1.7,1); 
\draw[->,> = stealth', shorten > = 1pt,mojemodra] (-1,1) -- (0,2); 

\draw[->,> = stealth', shorten > = 1pt,auto,decorate, decoration={snake, pre length=3pt,post length=3pt},mojecervena] (0,5) -- (-0.8,6);
\draw[->,> = stealth', shorten > = 1pt,auto,decorate, decoration={snake, pre length=3pt,post length=3pt},mojecervena] (1.7,5) -- (2.5,6);

\node at (0.5,4.5) {$|$ j $\rangle$};
\node at (0.5,3.5) {$|$ g $\rangle$};
\node at (0.5,2.5) {$|$ i $\rangle$};
\node at (0.5,1.5) {$|$ g $\rangle$};
\node at (0.5,0.5) {$|$ g $\rangle$};

\node at (1.2,4.5) {$\langle$ j $|$};
\node at (1.2,3.5) {$\langle$ j $|$};
\node at (1.2,2.5) {$\langle$ j $|$};
\node at (1.2,1.5) {$\langle$ j $|$};
\node at (1.2,0.5) {$\langle$ g $|$};

\node at (2.0,3.5) {$\tau$};
\node at (2.0,2.5) {$|T|$};
\node at (1.9,1.5) {$t$};
\node at (1.9,4.5) {$t'$};
\end{tikzpicture}
\end{tabular}}
\caption{Feynman diagrams for oscillatory contributions to F-2DES for $T>0$ and $T<0$.}
\label{FD_oscillations}
\end{figure}

In limit of short pulses, we can straight from Feynman diagrams in Fig. \ref{FD_oscillations} write the oscillatory signal. For real absorptive spectra, we can write
\begin{equation}
\begin{split}
&-Re\left(|\mu_i|^2|\mu_j|^2 \tilde{G}^*_{gj}(\omega_{t})e^{-\Gamma_{ij}|T|} e^{-i\omega_{ij}|T|} \tilde{G}^*_{gj}(\omega_{\tau})\right)\\
&-Re\left(|\mu_i|^2|\mu_j|^2 \tilde{G}^*_{gi}(\omega_{t})e^{-\Gamma_{ij}|T|} e^{i\omega_{ij}|T|} \tilde{G}^*_{gi}(\omega_{\tau})\right)\\
&-Re\left(|\mu_i|^2|\mu_j|^2 \tilde{G}^*_{gj}(\omega_{t})e^{-\Gamma_{ij}|T|} e^{-i\omega_{ij}|T|} \tilde{G}_{gi}(\omega_{\tau})\right)\\
&-Re\left(|\mu_i|^2|\mu_j|^2 \tilde{G}^*_{gi}(\omega_{t})e^{-\Gamma_{ij}|T|} e^{i\omega_{ij}|T|} \tilde{G}_{gj}(\omega_{\tau})\right)\\
\end{split}    
\end{equation}
The same could be done for opposite order of the pulses. Here, if we neglect difference between absorptive and excited-state stimulated-emission lineshapes, we get for $T>0$ the same real absorptive signal as was derived for $T>0$.
\begin{equation}
\begin{split}
&-Re\left(|\mu_i|^2|\mu_j|^2 \tilde{G}^*_{gj}(\omega_{t})e^{-\Gamma_{ij}|T|} e^{-i\omega_{ij}|T|} \tilde{G}^*_{gj}(\omega_{\tau})\right)\\
&-Re\left(|\mu_i|^2|\mu_j|^2 \tilde{G}^*_{gi}(\omega_{t})e^{-\Gamma_{ij}|T|} e^{i\omega_{ij}|T|} \tilde{G}^*_{gi}(\omega_{\tau})\right)\\
&-Re\left(|\mu_i|^2|\mu_j|^2 \tilde{G}^*_{gi}(\omega_{t})e^{-\Gamma_{ij}|T|} e^{i\omega_{ij}|T|} \tilde{G}_{gj}(\omega_{\tau})\right)\\
&-Re\left(|\mu_i|^2|\mu_j|^2 \tilde{G}^*_{gj}(\omega_{t})e^{-\Gamma_{ij}|T|} e^{-i\omega_{ij}|T|} \tilde{G}_{gi}(\omega_{\tau})\right)\\
\end{split}
\end{equation}
Clearly, the expressions for the positive and negative time are identical, and the oscillatory component is thus symmetric in F-2DES and F-PP. We thus expect the electronic coherence to be strongly suppressed in the difference signal, as we have indeed seen in the F-PP data of the 6-molecule aggregate and electronic dimer and in the F-2DES data of LH2, see Figs. (4,5,6) in the main article. 

\section{Pulse overlap region}
\begin{figure}
\setlength{\tabcolsep}{0mm} 
\def\arraystretch{0.0} 
\centering
\includegraphics[width=1.0\textwidth]{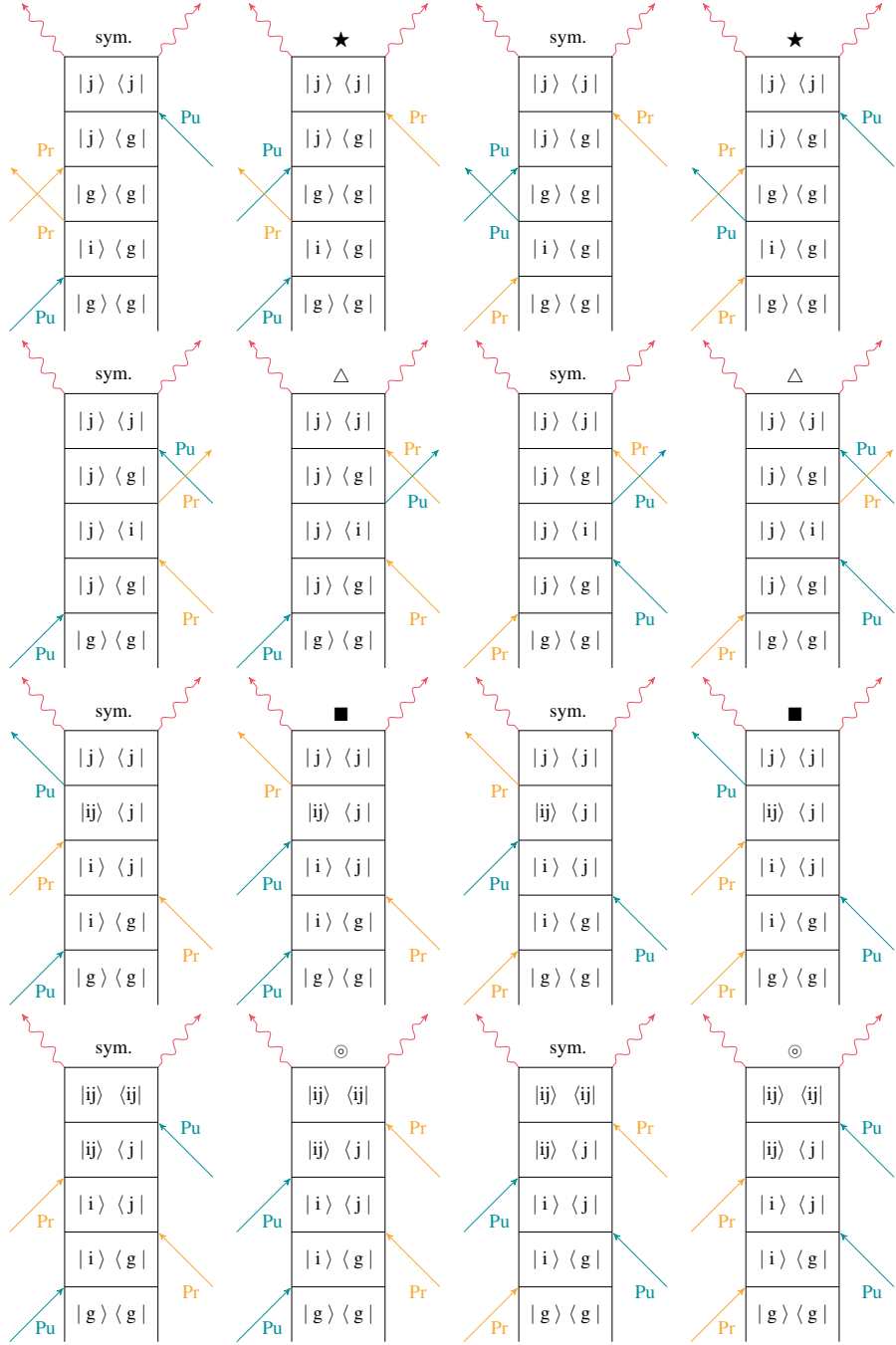}
\caption{Contributions to the signal for $T=0$\,fs.}
\label{pulseoverlappathways}
\end{figure}

We have already proved that for pathways with signature (1,-1,1,-1) we can subtract unwanted background and oscillations but at the same time dynamics remains, which leads to its emphasis. When we draw all possible Feynman diagrams with wrong pulse ordering, we can divide them into groups where the diagrams cancel each other after subtracting negative times. In figure \ref{pulseoverlappathways}, there are some pathways, noted as sym., that are symmetric for $T>0$ and $T<0$. Other diagrams are all in pairs, denoted with the same symbol ($\bigstar$,$\triangle$, $\blacksquare$, $\circledcirc$), which also have the same contribution, one in $T>0$ and the other in $T<0$. From this, it would follow that in the difference signal even pulse overlap artifacts could be removed. This was also supported by numerical simulations.   

\section{Numerical calculation of F-PP signal}

F-PP signal of six weakly coupled molecules with $15\;\textrm{fs}$ long pulses, presented in Fig. 4 in our article, was calculated with a custom script for the Quantarhei library. \cite{manvcal2020quantarhei} The calculation is basically the same as in our previous work.\cite{maly2021fluorescence} We numerically integrate the following master equation for the system density matrix $\rho(t')$:
\begin{equation}
    \frac{\partial \rho}{\partial t'} = -\frac{i}{\hbar}[H_s,\rho(t')] - \mathcal{R}\rho(t')+\frac{i}{\hbar}[\hat{\Vec{\mu}}\cdot\Vec{E},\rho(t')].
\label{mastereq}
\end{equation}
\noindent Here, $\mathcal{R}$ is relaxation a tensor in Lindblad form,\cite{valkunas2013molecular} $H_s$ is the Hamiltonian of the system, $\hat{\Vec{\mu}}$ is the transition dipole moment operator and $\Vec{E}$ is the electric field of the pump and probe pulses. For simplicity and to avoid orientational averaging, we take all transition dipoles parallel to the linear polarization of the electric field, with parallel pump and probe polarizations.

Solving equation \eqref{mastereq}, we obtain a time-dependent reduced density matrix $\rho(t',T,t)$, with which we can calculate the fluorescence signal using Eq.  \eqref{Fluorescence_signal_calculation}. To deal with the separation of timescales for the excitation and emission, we propagate the density matrix until time $t_{max}$ after the interaction with all three pulses and the system quasi-equilibration in its excited state, integrating the fluorescence signal. Fluorescence emitted after the time $t_{max}$ is computed analytically assuming an exponential decay of the excited-state population with the value at $t_{max}$ as an initial condition. To obtain the F-PP spectrum, we perform a discrete Fourier transform ($t \longrightarrow  \omega_{t}$) of the fluorescence signal and subtract the unpumped spectrum calculated separately without the pump pulse. Similarly to our previous work\cite{maly2021fluorescence}, we do not consider the Stokes shift in our calculations. 

In all simulations, the dynamics was calculated in the rotating wave approximation, using Taylor expansion of the propagator exponential into second order with time step in propagation $dt'=0.5$\,fs. The pulse delay $t$ between probes was scanned in steps $dt=4$\,fs. For time delay $T$ between pump and the first probe, we have set time step $dT=5$\,fs. For each set of pulse delays, we carried out 1x3x3 phase cycling\cite{tan2008theory} of the pump pulse $(\varphi_{1}=\varphi_2=0)$ and the probe pulse pair $(\varphi_{3}=l\cdot \frac{2\pi}{3},\varphi_4=m\cdot \frac{2\pi}{3})$. Further, we processed the fluorescence signal according to Eq. \eqref{phase_cycling}), isolating only the signal with phase signature $(\alpha,-\alpha,\beta,\gamma)$. 

\begin{equation}
   \mbox{FL}(\beta,\gamma)=\frac{1}{3\cdot 3}\sum_{m=0}^{2} \sum_{l=0}^{2} \mbox{FL}\left(l\cdot \frac{2\pi}{3} , m\cdot \frac{2\pi}{3}\right)e^{-il\cdot\beta\frac{2\pi}{3}}e^{-im\cdot\gamma\frac{2\pi}{3}} 
\label{phase_cycling}   
\end{equation}

In order to mimic the experimental condition used for measurement of some of the data (see Fig. 5 in the main article), we have used in the numerical simulation a partially rotating frame\cite{maly2021fluorescence} thus our second probe has additional phase $-(1-\gamma)\omega_0 t'$, where $\omega_0$ is the mean frequency of the pulse and $\gamma=0.2$. Further, it allows us to use longer time steps $dt$.

\bibliography{si_references}